\documentclass[showpacs,twocolumn]{revtex4}
\emergencystretch=\maxdimen
\usepackage{graphicx}
\usepackage{dcolumn}
\usepackage{amsmath}
\usepackage[latin1]{inputenc}
\usepackage{graphicx, psfrag}
\usepackage{amssymb}
\usepackage[colorlinks=true, citecolor=blue, urlcolor = blue, linkcolor= red, bookmarks=true]{hyperref}
\usepackage{float}
\usepackage{amsmath}
\usepackage{amsfonts}
\usepackage{dcolumn}
\usepackage{hyperref}
\usepackage{subfigure}
\usepackage{pgfplots}
\usepackage{epstopdf}
\usepackage{booktabs}

\makeatletter
\def\btt#1{\texttt{\@backslashchar#1}}
\DeclareRobustCommand\bblash{\btt{\@backslashchar}} \makeatother

\begin{document}

\title[]{Gravitational deflection of light and shadow cast by rotating Kalb-Ramond black holes}
\author{Rahul Kumar$^{a}$}\email{rahul.phy3@gmail.com}

\author{Sushant~G.~Ghosh$^{a,\;b}$} \email{sghosh2@jmi.ac.in, sgghosh@gmail.com}
\author{Anzhong Wang$^{c, d}$}\email{anzhong$\_$wang@baylor.edu}

\affiliation{$^{a}$ Centre for Theoretical Physics, Jamia Millia
	Islamia, New Delhi 110025, India}
\affiliation{$^{b}$ Astrophysics and Cosmology
	Research Unit, School of Mathematics, Statistics and Computer Science, University of
	KwaZulu-Natal, Private Bag 54001, Durban 4000, South Africa}
\affiliation{ $^{c}$ GCAP-CASPER, Physics Department, Baylor University, Waco, Texas 76798-7316, USA\\
	$^{d}$ Institute for Theoretical Physics and Cosmology, Zhejiang University of Technology, Hangzhou, 310032, China}

\date{\today}

\begin{abstract}
The nonminimal coupling of the nonzero vacuum expectation value of the self-interacting antisymmetric Kalb-Ramond field with gravity leads to a power-law hairy black hole having a parameter $s$,  which encompasses the Reissner$-$Nordstrom black hole ($s=1$). We obtain the axially symmetric counterpart of this hairy solution, namely, the rotating Kalb-Ramond black hole, which encompasses, as special cases, Kerr ($s=0$) and Kerr-Newman ($s=1$) black holes.  Interestingly, for a set of parameters ($M, a$, and $\Gamma$), there exists an extremal value of the Kalb-Ramond parameter ($s=s_{e}$), which corresponds to an extremal black hole with degenerate horizons, while for $s<s_{e}$, it describes a nonextremal black hole with Cauchy and event horizons, and no black hole for $s>s_{e}$. We find that the extremal value $s_e$ is also influenced by these parameters. The black hole shadow size decreases monotonically and the shape gets more distorted with an increasing $s$; in turn, shadows of rotating Kalb-Ramond black holes are smaller and more distorted than the corresponding Kerr black hole shadows. We investigate the effect of the Kalb-Ramond field on the rotating black hole spacetime geometry and analytically deduced corrections to the light deflection angle from the Kerr and Schwarzschild black hole values. The deflection angle for Sgr A* and the shadow caused by the supermassive black hole M87* are included and compared with analogous results of Kerr black holes. The inferred circularity deviation $\Delta C\leq 0.10$ for the M87* black hole merely constrains the Kalb-Ramond field parameter, whereas shadow angular diameter $\theta_d=42\pm 3\, \mu$as, within the $1\sigma$ region, places bounds $\Gamma\leq 0.09205$ for $s=1$ and  $\Gamma\leq 0.02178$ for $s=3$.
\end{abstract}

\maketitle
\section{Introduction}
The Kalb-Ramond field \cite{Kalb:1974yc} appears as a self-interacting second-rank antisymmetric tensor field in the heterotic string theory \cite{Gross:1984dd} and is attributed as the closed string excitation. The nonminimal coupling of the nonzero vacuum expectation value of the tensor field with the gravity sector leads to the spontaneous Lorentz symmetry violation: The ground state of a physical quantum system is characterized by nontrivial vacuum expectation values \cite{Altschul:2009ae,Kostelecky1}. It is found that the presence of the Kalb-Ramond field leads to many interesting implications, namely, the derived third-rank antisymmetric tensor can act as a spacetime torsion \cite{Majumdar:1999jd}, topological defects lead to the intrinsic angular momentum to the structures in galaxies \cite{Letelier:1995ze}, affect the observed anisotropy in the cosmic microwave background \cite{Maity:2004he}, provide crucial insights in the leptogenesis \cite{Ellis1}, and so on. The Kalb-Ramond field has been studied widely in the context of gravity and particle physics \cite{Seifert1,Chakraborty:2016lxo}. The compelling resemblance of the Kalb-Ramond field with the spacetime torsion ascertains that the Einstein gravity with the Kalb-Ramond field as a source is equivalent to a modified theory of gravity incorporating the spacetime torsion. The Solar System based tests, employed to test general relativity, reveal that the change incurred in the bending of light or perihelion precession of Mercury due to the presence of the Kalb-Ramond field would produce very tiny effects impossible to be detected with present-day precision \cite{Kar:2002xa}. However, the possibilities of detection in quasars or black hole spacetimes, where the spacetime curvature effects are strong, are still open and will have far-reaching consequences \cite{Banerjee:2017npv}. 

The Kalb-Ramond field can be considered as a generalization of the electromagnetic potential with two indices, such that the gauge potential $A_{\mu}$ is replaced by the second-rank antisymmetric tensor field $B_{\mu\nu}$ associated with the gauge-invariant rank-3 antisymmetric field strength $H_{\alpha\mu\nu}$, viz., $H_{\alpha\mu\nu}=\partial_{[\alpha}B_{\mu\nu]}$; $H_{\alpha\mu\nu}$ is analogous to the Faraday field tensor $F_{\mu\nu}$ \cite{Kalb:1974yc}. The Einstein-Hilbert action nonminimally coupled with the self-interacting Kalb-Ramond field 
reads \cite{Altschul:2009ae}
\begin{align}
S&=\int \sqrt{-g}d^4x\Big(\frac{R}{16\pi G}-\frac{1}{12}H_{\alpha\mu\nu}H^{\alpha\mu\nu}-V(B_{\mu\nu}B^{\mu\nu}\nonumber\\
&\pm b_{\mu\nu}b^{\mu\nu})+\frac{1}{16\pi G}\left(\xi_2 B^{\mu\lambda}B^{\nu}_{\lambda}R_{\mu\nu}+\xi_3B_{\mu\nu}B^{\mu\nu}R\right) \Big),\label{action}
\end{align}
where $R$ and $R_{\mu\nu}$ are, respectively, the Ricci scalar and Ricci tensor and $\xi_{2,3}$ are the nonminimal coupling constants. The potential term $V$ drives the development of a nonzero vacuum expectation value for the tensor field, i.e., $\langle B_{\mu\nu}\rangle=b_{\mu\nu}$, which breaks local Lorentz and diffeomorphism symmetry. The static spherically symmetric solution of the modified Einstein equations leads to the hairy black hole solution \cite{Bessa:2019pom}
\begin{eqnarray}
ds^2&=&-\left(1-\frac{2M}{r}+\frac{\Gamma}{r^{2/s}}\right)dt^2+\frac{1}{\left(1-\frac{2M}{r}+\frac{\Gamma}{r^{2/s}}\right)}dr^2\nonumber\\
&&+r^2d\theta^2+r^2\sin^2\theta d\phi^2.\label{NR}
\end{eqnarray}
Here, $M$ is the black hole mass, and $\Gamma$ and $s$ are the spontaneous Lorentz violating parameters related to the vacuum expectation value of the Kalb-Ramond field and the nonminimal coupling parameters, viz., $s=|b^2|\xi_2$ with $b^2=b_{\mu\nu}b^{\mu\nu}$. $M$ has dimension of length $[L]$, whereas $\Gamma$ has dimensions of $[L]^{2/s}$. The power-law hairy black hole (\ref{NR}) encompasses the Schwarzschild solution when $s= 0$, and Schwarzschild de-Sitter for $s=-1$, and when $s=1$ it resembles the Reissner$-$Nordstrom black hole. However, nonrotating black holes can hardly be tested by observations, as black hole spin is crucial for the astrophysical processes. The Kerr metric \cite{Kerr:1963ud} is one of the most important solutions of general relativity which represents a rotating black hole that can result from gravitational collapses.  This prompted us to seek an axisymmetric generalization of the metric (\ref{NR}) or finding a Kerr-like metric, namely, a rotating Kalb-Ramond black hole metric, and to test it with astrophysical observations. We discuss the various black hole properties including the horizon structure and the static limit surfaces, calculate the corresponding conserved quantities, and establish the Smarr formula. We explore the Kalb-Ramond field signatures in black hole spacetimes in the context of available astrophysical observations. Then, we study photon motion in the rotating Kalb-Ramond black hole spacetime, as they play crucial roles in determining the strong gravitational field features, such as gravitational lensing and shadow. Furthermore, the Gauss-Bonnet theorem is utilized to discuss the gravitational lensing of light and to analytically calculate the deflection angle in the weak-field limit caused by the rotating Kalb-Ramond black hole, considering the source and observer at finite distances from the black hole. The correction in the deflection angle due to the presence of the Kalb-Ramond field for the supermassive black hole Sgr A* at the Galactic center is estimated and found to be within the resolution of today's observational facilities. Moreover, the recent observation of the M87* black hole shadow by the Event Horizon Telescope (EHT) Collaboration has facilitated direct probing of the near horizon regime and offers an unprecedented opportunity to test the nature of strong gravity \cite{Akiyama:2019cqa,Akiyama:2019eap,Akiyama:2019fyp,Akiyama:2019bqs}. We examine the viability of the obtained rotating black hole in attributing the observed asymmetry in the M87* black hole emission ring.

The organization of this paper is as follows. We begin in Sec.~\ref{sec2} with the construction of the rotating counterpart of the metric (\ref{NR}), namely, the rotating Kalb-Ramond metric. We also discuss generic features of the black hole including horizon structures and static limit surfaces. In Sec.~\ref{sec3}, we exploit the spacetime isometries to deduce the conserved mass and angular momentum of the rotating Kalb-Ramond black hole. The discussion of a black hole shadow and the effect of the Kalb-Ramond field on the shape and size of shadows are the subjects of Sec.~\ref{sec4}. In Sec.~\ref{sec5}, we set the premises for the gravitational deflection of light in the stationary spacetime and estimate the correction in the deflection angle owing to the Kalb-Ramond field. Finally, we summarize our main findings in  Sec.~\ref{sec6}.\\

\section{Rotating black hole}\label{sec2}
Here, we find the stationary and axisymmetric counterpart of the spherically symmetric solution (\ref{NR}) governed by four parameters $M$, $\Gamma$, $a$, and a free parameter $s$ (Kalb-Ramond parameter) that measures the potential deviation from the Kerr solution \cite{Kerr:1963ud} and also generalizes the Kerr-Newman solution \cite{Newman:1965my}, which in Boyer-Lindquist coordinates reads 
\begin{align}
ds^2=&-\left(\frac{\Delta - a^2 \sin^2 \theta}{\Sigma}\right) dt^2 + \frac{ \Sigma}{\Delta }  \, dr^2  \nonumber \\
& -  2 a \sin^2 \theta \left(1 - \frac{\Delta - a^2 \sin^2 \theta}{\Sigma} \right) dt \, d \phi
+ \Sigma \, d \theta^2
\nonumber \\
&  +  \, \sin ^2 \theta  \left[ \Sigma + a^2 \sin^2 \theta \left(2 - \frac{\Delta - a^2 \sin^2\theta}{\Sigma}\right)   \right]    d \phi^2,\label{rotbhtr}
\end{align}
with 
\begin{equation}
\Delta=r^2+a^2-2Mr+\frac{\Gamma}{r^{-2(s-1)/s}},\quad \Sigma=r^2+a^2\cos^2\theta,
\end{equation} 
and $a$ is the spin parameter. The metric Eq.~(\ref{rotbhtr}) reverts to Kerr black holes as the special case  $s\to 0$, to Kerr-Newman black holes for $s=1$, and to spherically symmetric black holes (\ref{NR}) when only $a=0$. For definiteness, we call the four parameter metrics~(\ref{rotbhtr}) $-$ the rotating Kalb-Ramond black holes, which contain all known stationary black holes of general relativity. Interestingly, like the Kerr spacetime, the rotating Kalb-Ramond black hole spacetime metric (\ref{rotbhtr}) still possesses the time-translational and rotational invariance isometries, which, respectively, entail the existence of two Killing vector fields $\eta^{\mu}_{(t)}=\left(\frac{\partial}{\partial t}\right)^{\mu} $ and $\eta^{\mu}_{(\phi)}=\left(\frac{\partial}{\partial \phi}\right)^{\mu}$.

The event horizon is a null stationary surface that represents the locus of outgoing future-directed null geodesic rays that never manage to reach arbitrarily large distances from the black hole \cite{Hawking:1971vc,he,Poisson:2009pwt}. The outward normal to such surfaces is proportional to $\partial_{\mu}r$; therefore, horizons are defined by the surfaces $g^{\mu\nu}\partial_{\mu}r\partial_{\nu}r=g^{rr}=\Delta=0,$ and thus, their radii are zeros of
\begin{equation}
r^2+a^2-2Mr+\frac{\Gamma}{r^{-2(s-1)/s}}=0.\label{horizon}
\end{equation}
\begin{figure*}
	\begin{tabular}{c c}
		\includegraphics[scale=0.72]{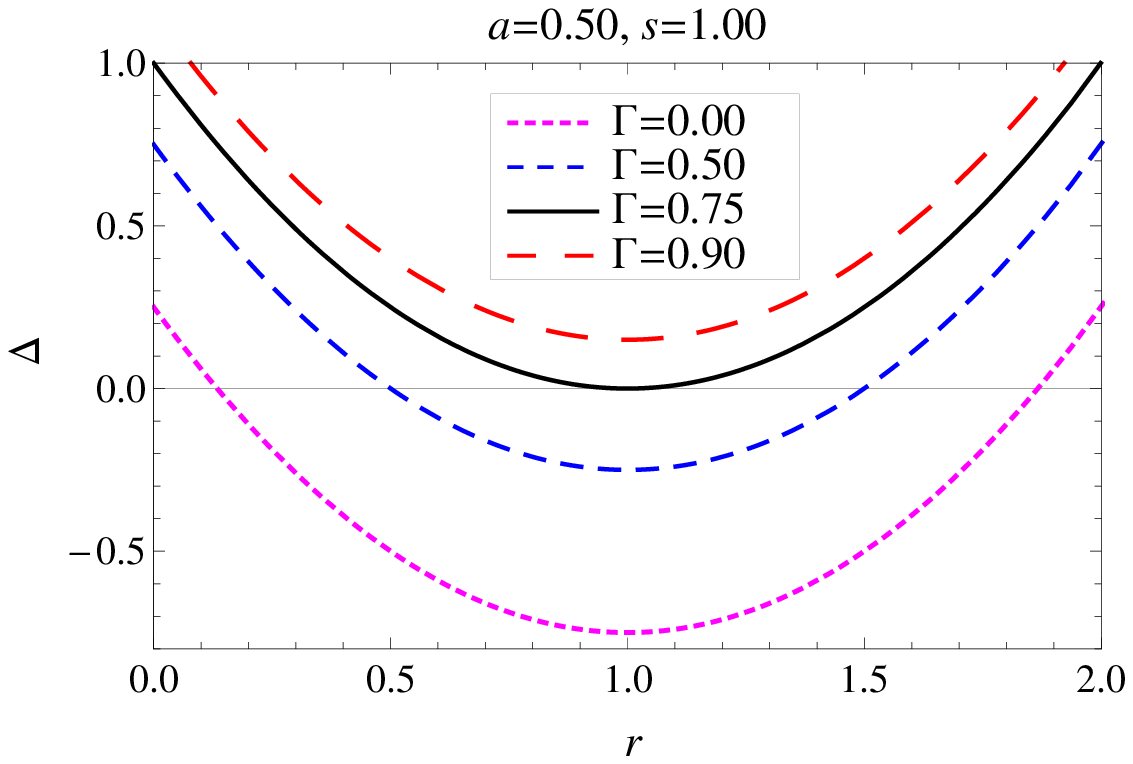}&
		\includegraphics[scale=0.72]{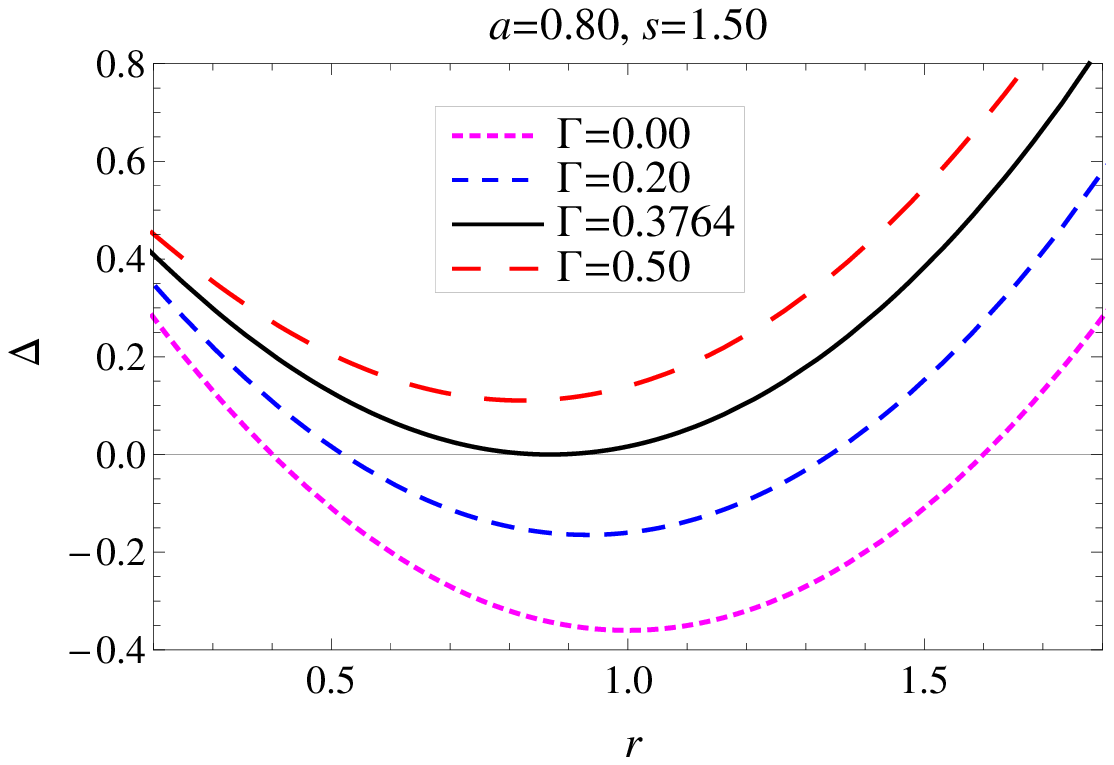}\\
		\includegraphics[scale=0.72]{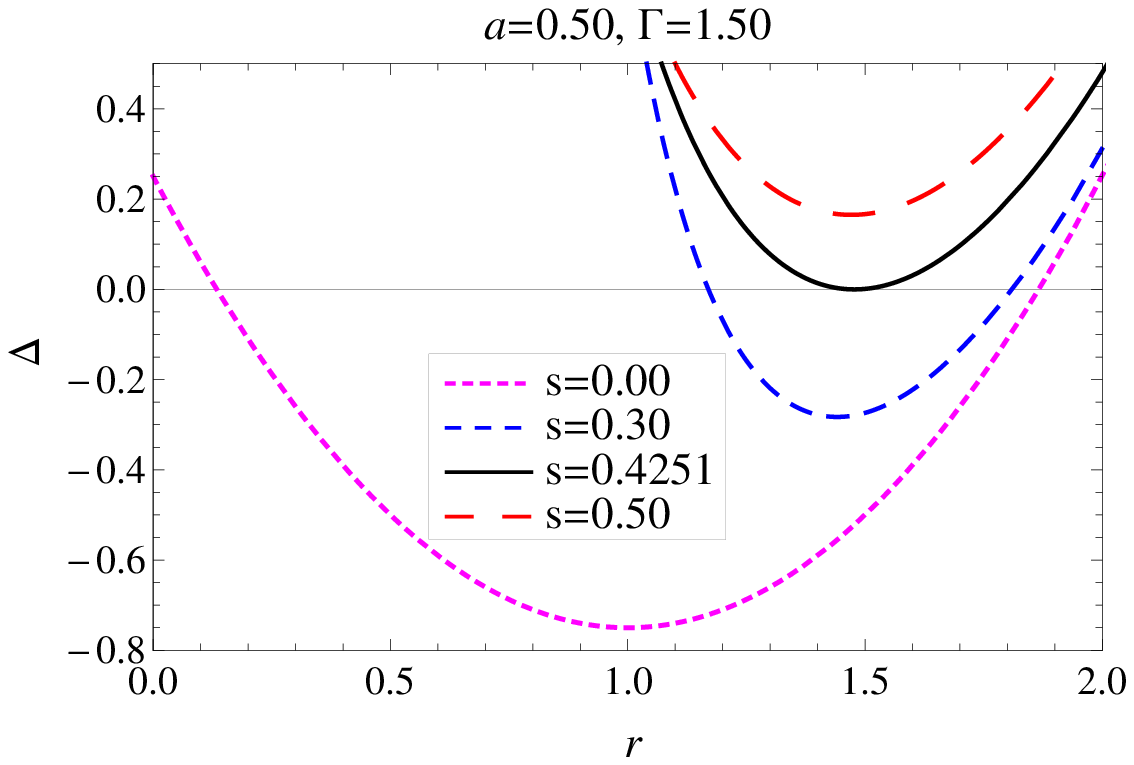}&
		\includegraphics[scale=0.72]{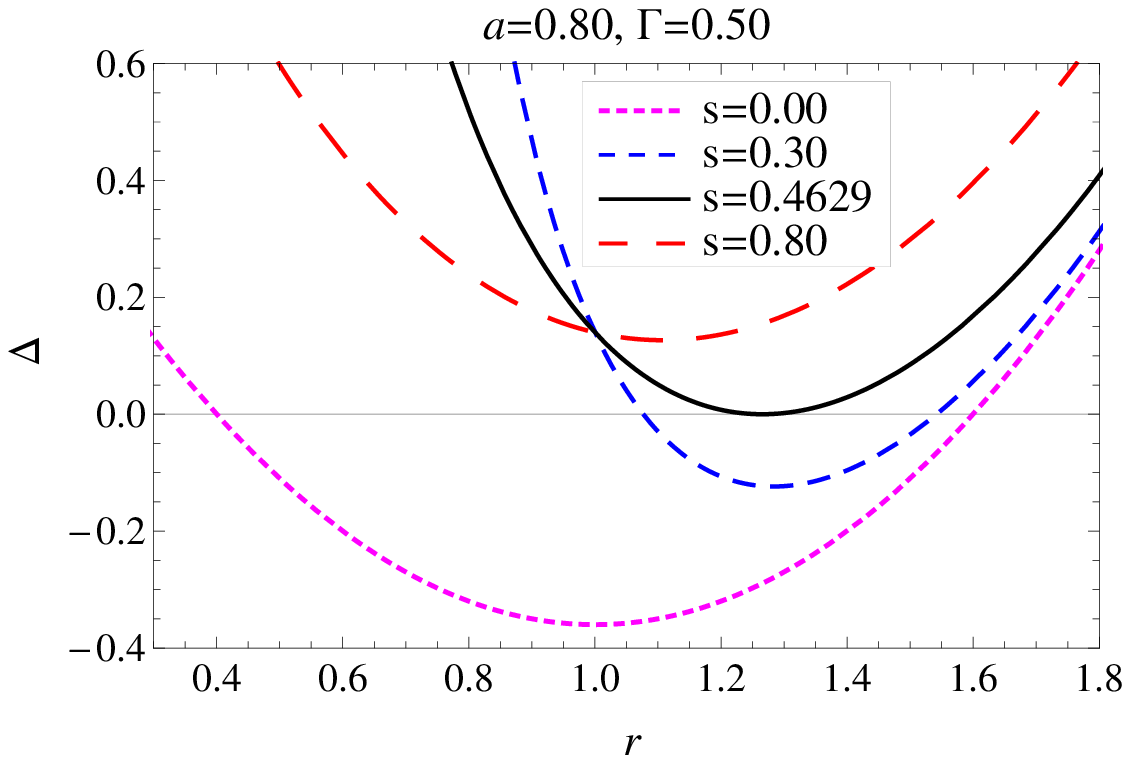}\\
		\includegraphics[scale=0.72]{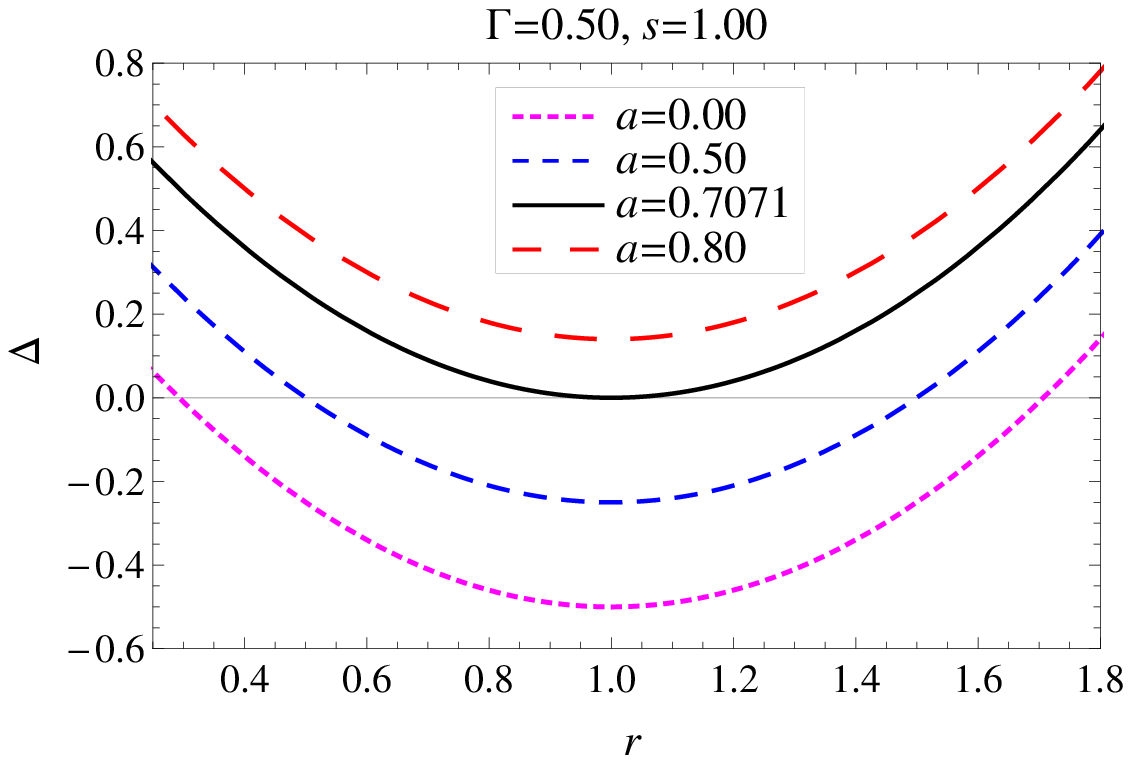}&
		\includegraphics[scale=0.72]{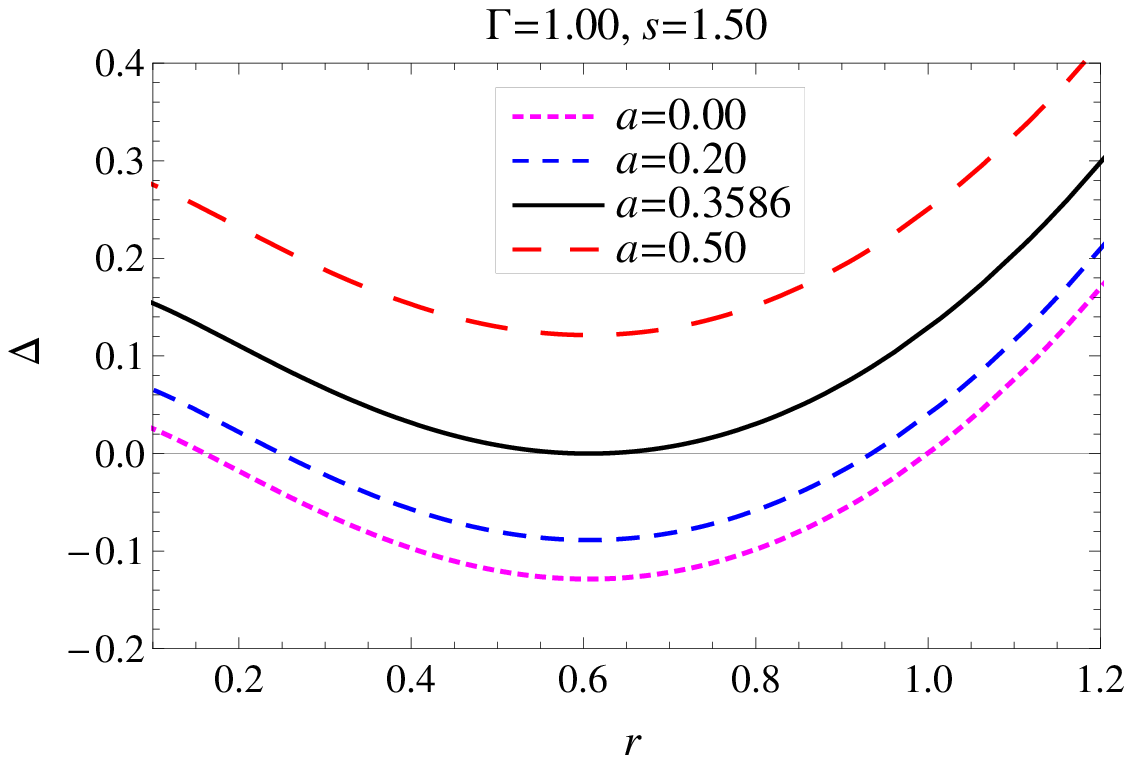}
	\end{tabular}
	\caption{The behavior of horizons with varying black hole parameters $a, \Gamma$, and $s$. The black solid line corresponds to the extremal black hole with degenerate horizons.}
	\label{Horizonfig}
\end{figure*}
For the special case $s=1$, Eq.~(\ref{horizon}) reduces to
\begin{equation}
r^2+a^2-2Mr+Q^2=0,\label{horizonKN}
\end{equation}
where $\Gamma$ is identified as the charge $Q^2$, and solutions of the above equation give radii of horizons for the Kerr-Newman black hole given by
\begin{equation}
r_{\pm}=M\pm\sqrt{M^2-a^2-Q^2}.
\end{equation} 
A numerical analysis of Eq.~(\ref{horizon}) reveals that it has maximum two real positive roots, corresponding to the inner Cauchy horizon ($r_-$) and outer event horizon ($r_+$), such that $r_-\leq r_+$ (cf. Fig.~ \ref{Horizonfig}). Two distinct real positive roots of $\Delta=0$ infers the nonextremal black hole, while no black hole in the absence of real positive roots of Eq.~(\ref{horizon}), i.e., no horizon exists. There exists a particular value of the parameter $s$, $s=s_e$, for which an extremal black hole occurs, such that Eq.~(\ref{horizon}) admits a double root; i.e., the two horizons coincide $r_-=r_+=r_e$. We have explicitly shown that, for fixed values of $a$ and $\Gamma$, $r_+$ decreases and $r_-$ increases with increasing $s$ and eventually coincide for the extremal value of $s$, i.e., $r_-=r_+=r_e$ for $s=s_e$ (cf. Fig.~ \ref{Horizonfig}). Horizon radii vary in a similar way with increasing $a$ and $\Gamma$. Moreover, the numerical analysis infers that it is possible to find extremal values of parameters $a=a_e$ for fixed $s$ and $\Gamma$, and $\Gamma=\Gamma_e$ for fixed $a$ and $s$, for which algebraic equation $\Delta=0$ has double roots as depicted in Fig.~ \ref{Horizonfig}. Figure~\ref{Horizonfig} also shows that, for the fixed values of $M$ and $a$, the event horizon radii for rotating Kalb-Ramond black holes are smaller as compared to those for the Kerr black hole, which is a potentially generic effect of certain classes of  gravity theories \cite{Held:2019xde}.
\begin{figure*}
	\begin{tabular}{c c}
		\includegraphics[scale=0.72]{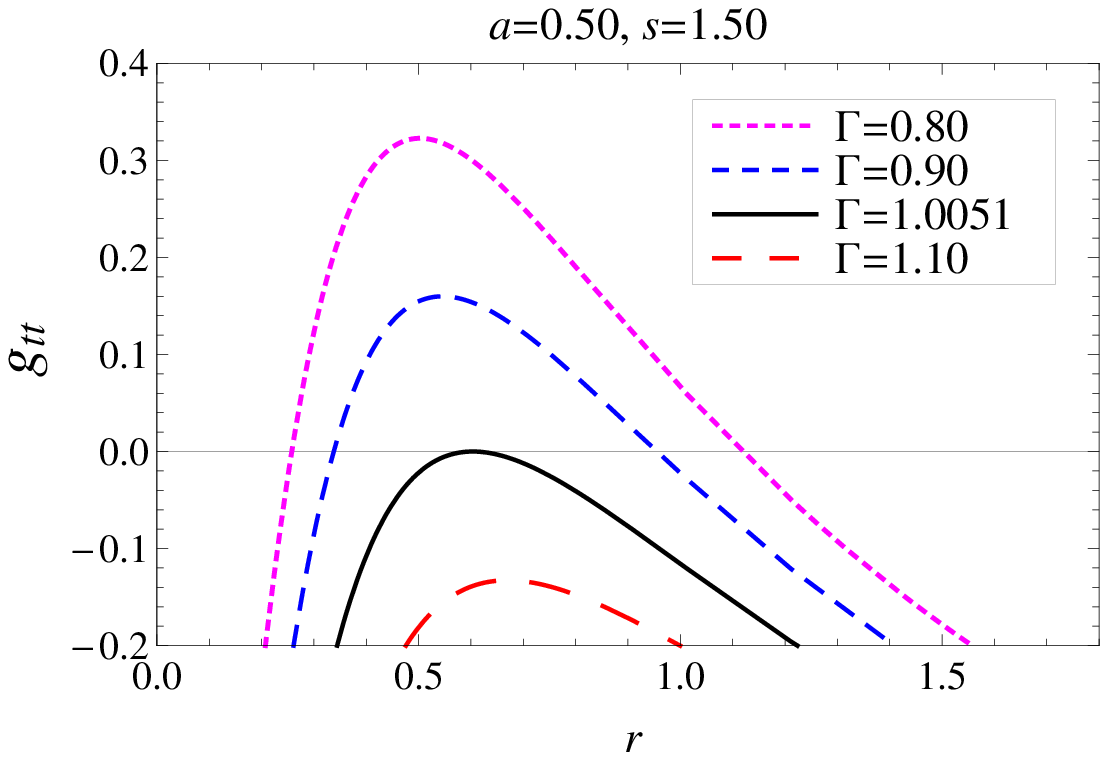}&
		\includegraphics[scale=0.72]{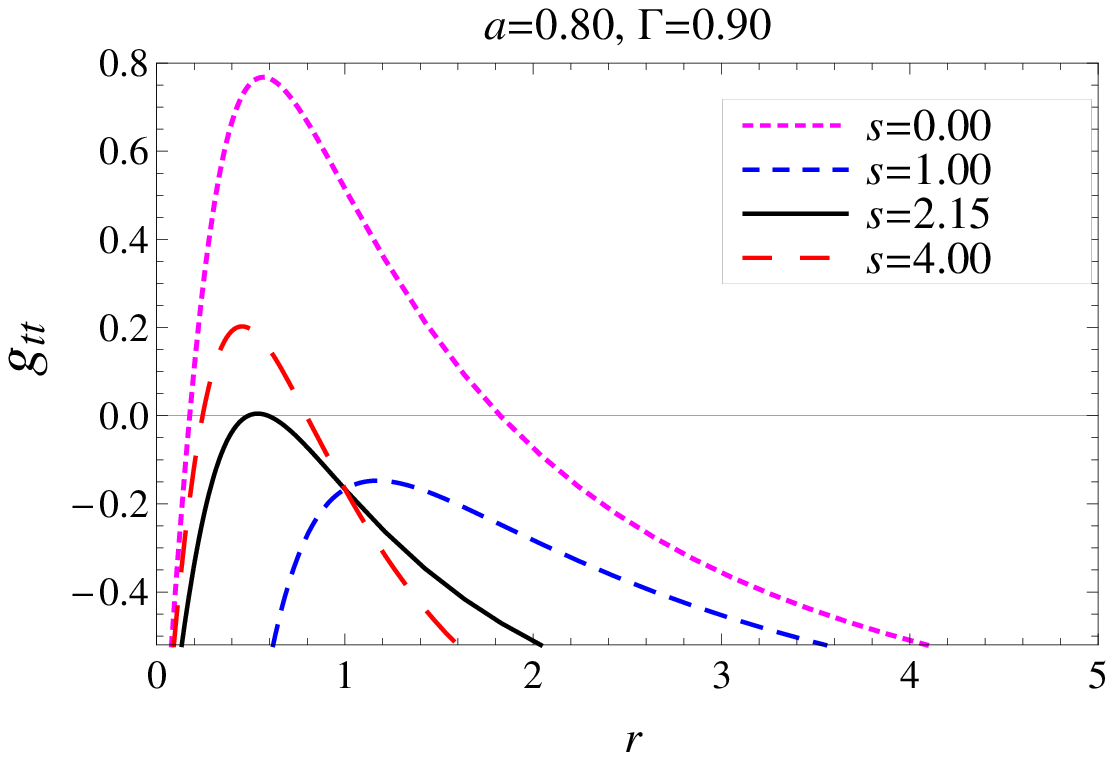}\\
		\includegraphics[scale=0.72]{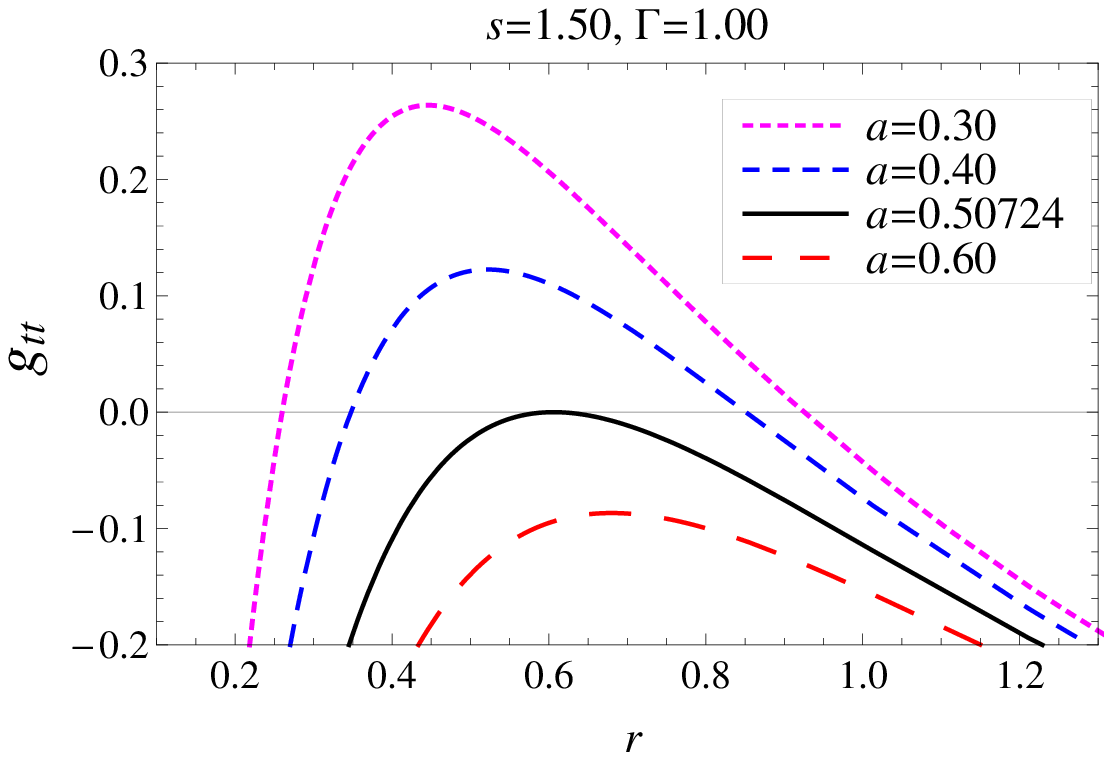}&
		\includegraphics[scale=0.72]{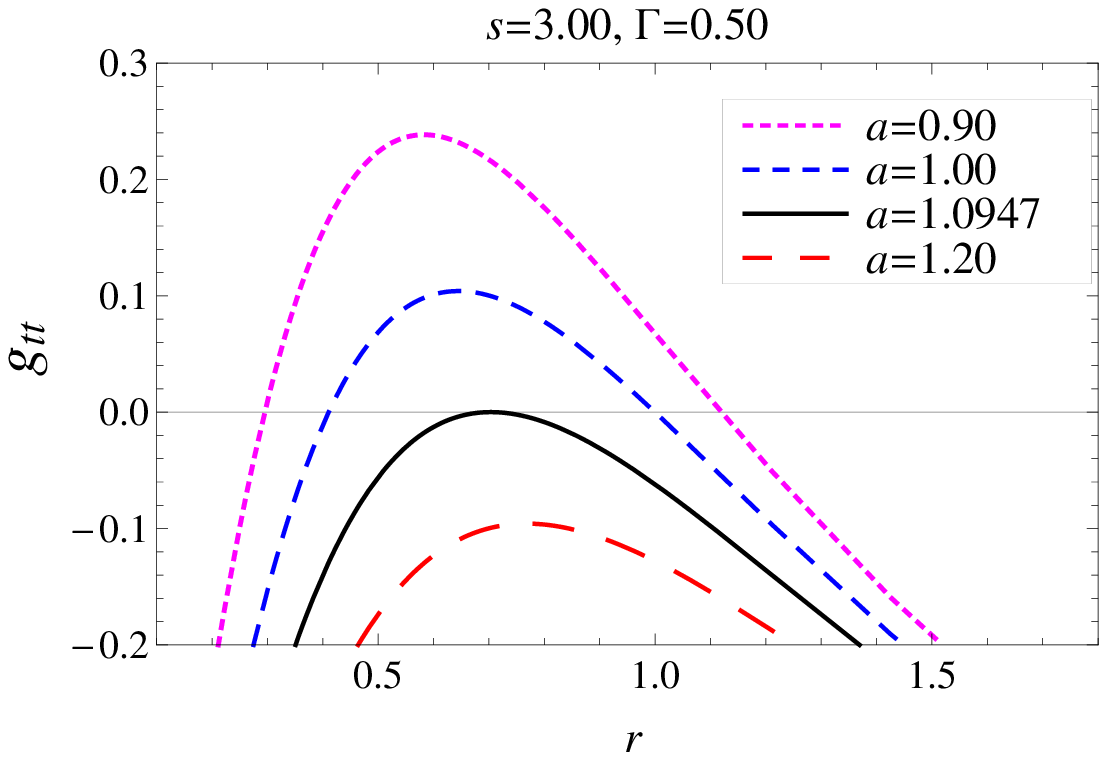}
	\end{tabular}
	\caption{The behavior of the SLS with varying parameters $a, \Gamma$, $s$ and $\theta=\pi/4$. The black solid curve in each plot corresponds to the degenerate SLS.} \label{gtt}
\end{figure*}

The static observers in the stationary spacetime follow the worldline of timelike Killing vector $\eta^{\mu}_{(t)}$, such that their four-velocity is $u^{\mu}\propto \eta^{\mu}_{(t)}$ with the proper normalization factor. These observers can exist as long as $\eta^{\mu}_{(t)}$ is timelike, such that $\eta^{\mu}_{(t)}\eta_{\mu(t)}=g_{tt}=0$ or
\begin{equation}
r^2+a^2\cos^2\theta- 2Mr+\frac{\Gamma}{r^{-2(s-1)/s}}=0\label{gtteq}
\end{equation}
defines the boundary of the static limit surface (SLS), which apart from black hole parameters also depends on $\theta$ and coincides with the event horizon only at the poles. For the particular case $s=1$, Eq.~(\ref{gtteq}) corresponds to the Kerr-Newman black hole as
\begin{equation}
r^2+a^2\cos^2\theta- 2Mr+Q^2=0\label{gtteqKN}
\end{equation}
and admits the solutions
$$r_{SLS}^{\pm}=M\pm\sqrt{M^2-a^2\cos^2\theta-Q^2},$$
which can be identified as the SLS radii for the Kerr-Newman black hole. Equation (\ref{gtteq}) is solved numerically and the behavior of SLS is shown in Fig.~\ref{gtt}. It is clear from Fig.~\ref{gtt} that radii of the SLS decrease with increasing $\Gamma$ and $a$. The two SLS, corresponding to the real positive roots of Eq.~(\ref{gtteq}), coincide for suitably chosen parameters. However, these extremal values are different from those for the degenerate horizons. For fixed values of $M$ and $a$, the SLS radii for the rotating Kalb-Ramond black holes are smaller than the Kerr black hole values. Likewise the Kerr black hole, apart from $\Delta=0$, which is merely a coordinate singularity, rotating metric (\ref{rotbhtr}) is also singular at $\Sigma=0$, which is attributed to a ring-shaped physical singularity at the equatorial plane of the center of the black hole with radius $a$.

Zero angular momentum observers are the stationary observers with zero angular momentum with respect to spatial infinity, but due to frame dragging they have the position-dependent angular velocity $\omega$:
\begin{equation}
\omega=\frac{d\phi}{dt}=-\frac{g_{t\phi}}{g_{\phi\phi}}=\frac{2Mar-\frac{\Gamma a}{r^{-2(s-1)/s}}}{\left[(r^2+a^2)^2-a^2\Delta\right]},
\end{equation}
which increases as the observer approaches the black hole and eventually takes the maximum value at the event horizon:
\begin{equation}
\Omega=\left.\omega\right|_{r=r_+}=\frac{2Ma r_+-\frac{\Gamma a}{r_+^{-2(s-1)/s}}}{(r_+^2+a^2)^2}\label{angvelocity},
\end{equation}
such that observers are in a state of corotation with the black hole. Here, $\Omega$ is the black hole angular velocity, which in the limits $s=0$ reads
\begin{equation}
\Omega=\frac{a}{r_+^2+a^2},
\end{equation}
and corresponds to the Kerr black hole value \cite{Poisson:2009pwt,Chandrasekhar:1992}.

\section{Komar Mass and Angular Momentum}\label{sec3}
The mass and angular momentum attributed to the stationary, asymptotically flat black hole spacetime correspond to the conserved quantities associated with the asymptotically timelike and spacelike Killing vectors fields, respectively, $\eta^{\mu}_{(t)} $ and $\eta^{\mu}_{(\phi)}$. A general argument for equality of the conserved Arnowitt-Deser-Misner mass \cite{Arnowitt:1962hi} and of the Komar mass \cite{Komar:1958wp} for stationary spacetimes having a timelike Killing vector is established in Refs.~\cite{Jaramillo:2010ay,Shibata:2004qz}. Following the Komar \cite{Komar:1958wp} definitions of conserved quantities, we consider a spacelike hypersurface $\Sigma_t$, extending from the event horizon to spatial infinity, which is a surface of constant $t$ with unit normal vector $n_{\mu}$  \cite{Chandrasekhar:1992,Wald}. The two-boundary $S_t$ of the hypersurface $\Sigma_t$ is a constant $t$ and constant $r$ surface with unit outward normal vector $\sigma_{\mu}$. The effective mass reads \cite{Komar:1958wp}
\begin{equation}
M_{\text{eff}}=-\frac{1}{8\pi}\int_{S_t}\nabla^{\mu}\eta^{\nu}_{(t)}dS_{\mu\nu},\label{mass}
\end{equation}
where $dS_{\mu\nu}=-2n_{[\mu}\sigma_{\nu]}\sqrt{h}d^2\theta$ is the surface element of $S_t$, $h$ is the determinant of ($2\times 2$) metric on $S_t$, and 
\begin{equation}
n_{\mu}=-\frac{\delta^{t}_{\mu}}{|g^{tt}|^{1/2}},\qquad \sigma_{\mu}=\frac{\delta^{r}_{\mu}}{|g^{rr}|^{1/2}},
\end{equation}
are, respectively, timelike and spacelike unit outward normal vectors. Thus, mass integral Eq.~(\ref{mass}) turned into an integral over closed 2-surface at infinity:
\begin{align}
M_{\text{eff}}=&\frac{1}{4\pi}\int_{0}^{2\phi}\int_{0}^{\phi}\frac{\sqrt{g_{\theta\theta}g_{\phi\phi}}}{|g^{tt}g^{rr}|^{1/2}}\nabla^{t}\eta^{r}_{(t)}d\theta d\phi\nonumber\\
=& \frac{1}{4\pi}\int_{0}^{2\phi}\int_{0}^{\phi}\frac{\sqrt{g_{\theta\theta}g_{\phi\phi}}}{|g^{tt}g^{rr}|^{1/2}}\left(g^{tt}\Gamma^{r}_{tt}+g^{t\phi}\Gamma^{r}_{t\phi} \right)d\theta d\phi.
\end{align}
Using the metric elements Eq.~(\ref{rotbhtr}), we obtain the effective mass of the rotating Kalb-Ramond black hole:
\begin{equation}
M_{\text{eff}}=M+\frac{1}{2ras}\left( (r^2+a^2)(s-2)\tan^{-1}\left(\frac{a}{r}\right)-a rs\right)\frac{\Gamma}{r^{[-(s-2)/s]}},\label{mass1}
\end{equation}
which is clearly corrected due to the Kalb-Ramond field, and goes over to the Kerr black hole case that is $M_{\text{eff}}=M$, when $s=0$. For the special case $s=1$, Eq.~(\ref{mass1}) resembles the effective mass for the Kerr-Newman black hole with $\Gamma$ as the electric charge $Q^2$ and reads \cite{Modak:2010fn}
\begin{equation}
M_{\text{eff}}=M-\frac{Q^2}{2r^2a}\left( (r^2+a^2)\tan^{-1}\left(\frac{a}{r}\right)+a r\right).
\end{equation}
The effective mass for the spherically symmetric Kalb-Ramond black hole ($a=0$) is obtained from Eq.~(\ref{mass1}) and reads
$$M_{\text{eff}}=M-\frac{1}{s}\frac{\Gamma}{r^{[-(s-2)/s]}},$$
which reduces to the Reissner$-$Nordstrom black hole for  $s=1$:
$$M_{\text{eff}}=M-\frac{Q^2}{r},$$
and to Schwarzschild black hole $M_{\text{eff}}=M$, when $s=0$.
\begin{figure*}
	 \begin{tabular}{c c}
\includegraphics[scale=0.72]{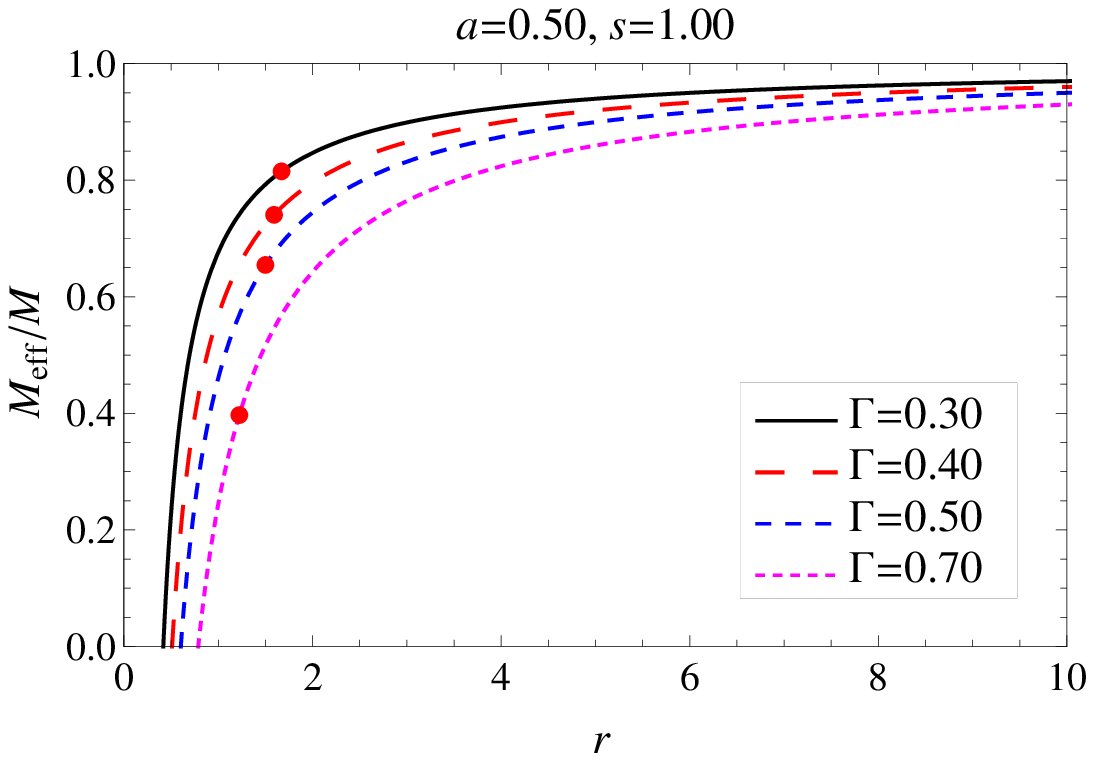}&
\includegraphics[scale=0.72]{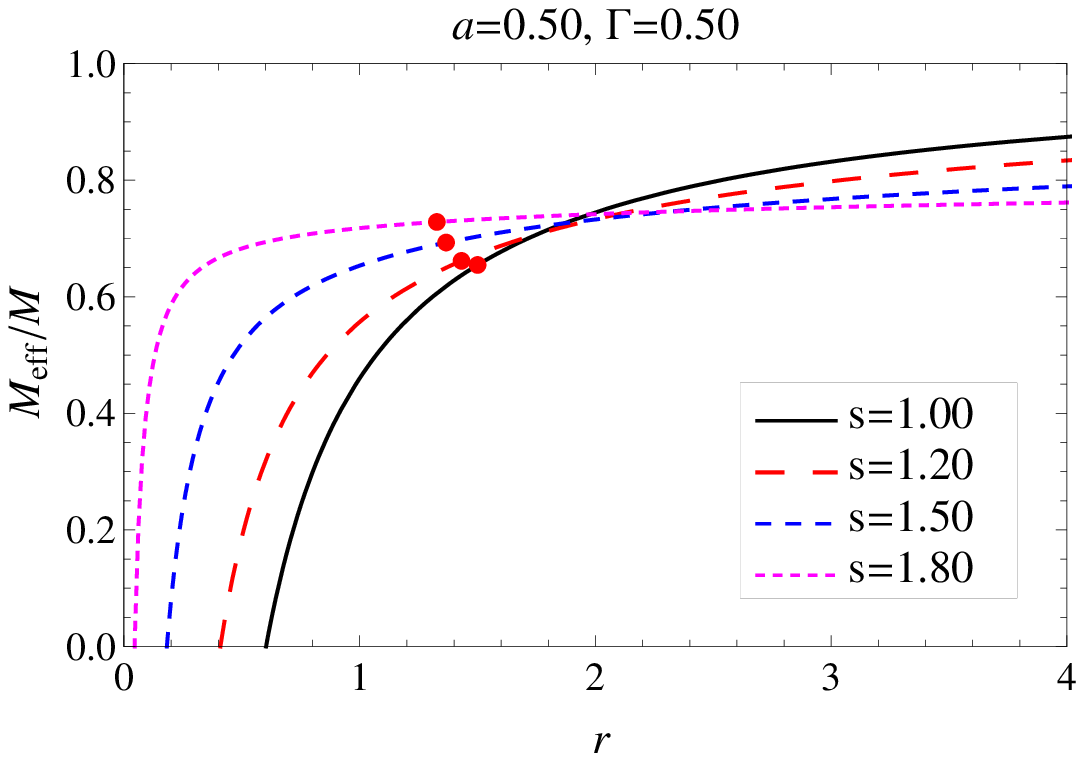}\\
\includegraphics[scale=0.72]{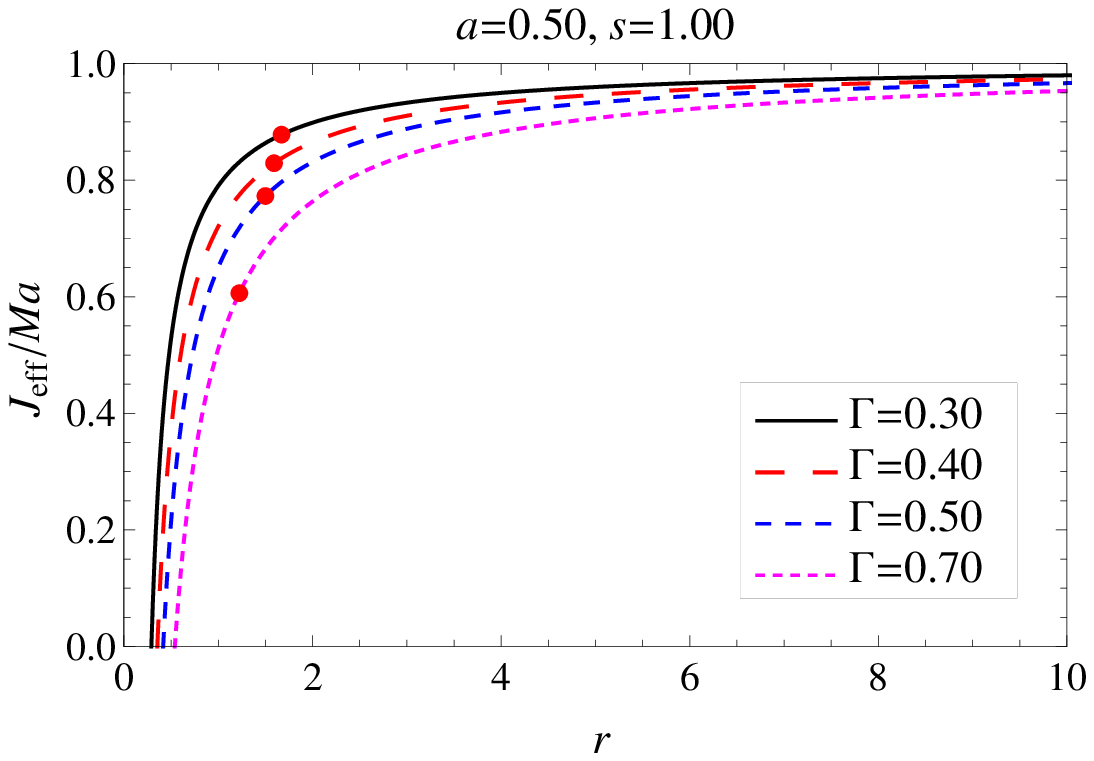}&
\includegraphics[scale=0.72]{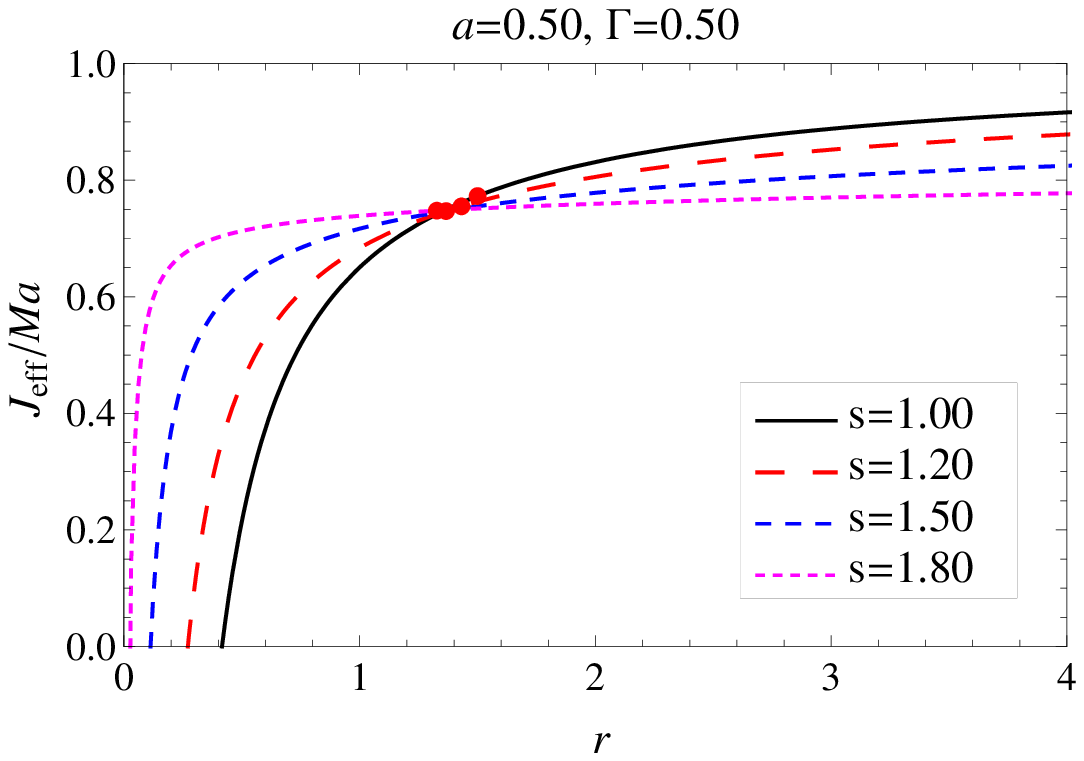}
     \end{tabular}
 	\caption{The behavior of effective mass and angular momentum vs $r$ for different values of the parameters. Black solid curves correspond to the Kerr-Newman black hole and red dots in each curve denote the locations of the event horizon. }
 \label{Komar}
\end{figure*}
Now, we use the spacelike Killing vector $\eta^{\mu}_{(\phi)}$ to calculate the effective angular momentum  \cite{Komar:1958wp}
\begin{equation}
J_{\text{eff}}=\frac{1}{16\pi}\int_{S_t}\nabla^{\mu}\eta^{\nu}_{(\phi)}dS_{\mu\nu},\label{ang}
\end{equation}
using the definitions of the surface element, Eq.~(\ref{ang}) recast as
\begin{align}
J_{\text{eff}}=&-\frac{1}{8\pi}\int_{0}^{2\phi}\int_{0}^{\phi}\nabla^{\mu}\eta^{\nu}_{(t)}n_{\mu}\sigma_{\nu}\sqrt{h}d\theta d\phi\nonumber\\
=& \frac{1}{8\pi}\int_{0}^{2\phi}\int_{0}^{\phi}\frac{\sqrt{g_{\theta\theta}g_{\phi\phi}}}{|g^{tt}g^{rr}|^{1/2}}\left(g^{tt}\Gamma^{r}_{t\phi}+g^{t\phi}\Gamma^{r}_{\phi\phi} \right)d\theta d\phi.
\end{align}
After performing the integration for the rotating Kalb-Ramond black hole Eq.~(\ref{rotbhtr}), this reads
\begin{eqnarray}
J_{\text{eff}}&=&Ma+\frac{\Gamma}{4ra^2s}\frac{1}{r^{[-(s-2)/s]}}\Big((r^2+a^2)^2(s-2)\tan^{-1}\left(\frac{a}{r}\right)\nonumber\\
&&-\left((3s-2)a^2+r^2(s-2)\right)ra\Big),\label{ang1}
\end{eqnarray}
which identically vanishes in the limiting case of $a=0$, and for the particular case of $s=1$ it reduces to
\begin{equation}
J_{\text{eff}}=Ma+\frac{\Gamma(r^2-a^2)}{4ar}-\frac{\Gamma}{4a^2r^2}(r^2+a^2)^2\tan^{-1}\left(\frac{a}{r}\right),
\end{equation}
which can be identified as the Kerr-Newman black hole value \cite{Modak:2010fn}. In the asymptotic limits $r\to\infty$, the effective angular momentum Eq.~(\ref{ang1}) restores the value $J_{\text{eff}}=Ma$, which corresponds to the value for the Kerr black hole. Thus, the effects of the Kalb-Ramond field subside at a very large distance from the black hole. Equations (\ref{mass1}) and (\ref{ang1}) imply that at a finite radial distance the values of the effective mass and angular momentum get modified from their asymptotic values and depend on the sign of $\Gamma$.
In Fig.~\ref{Komar}, we have shown the normalized effective mass and angular momentum variation with radial distance $r$ for various values of black hole parameters, such that, at asymptotically large $r$, the normalized values become unity, as expected. It is clear that, for fixed values of $a$ and $s$, the effective values of $M_{\text{eff}}/M$ and $J_{\text{eff}}/Ma$ decrease with increasing field parameter $\Gamma$, whereas, for fixed values of $a$ and $\Gamma$, the effective mass and angular momentum show diverse behavior with varying $s$.  Moreover, outside the event horizon, the effective angular momentum of the black hole reduces with increasing Kalb-Ramond field parameter $s$. Thus, for the rotating Kalb-Ramond black hole, the values of effective mass and effective angular momentum are smaller as compared to those for the Kerr black hole.

It is well known that the Killing vectors $\eta^{\mu}_{(t)}$ or $\eta^{\mu}_{(\phi)}$ are not the  generators of the stationary black hole horizon; rather, it is their specific linear combination \cite{Chandrasekhar:1992} as
\begin{equation}
\chi^{\mu}=\eta^{\mu}_{(t)}+\Omega \eta^{\mu}_{(\phi)},
\end{equation}
such that $\chi^{\mu}$ is globally timelike outside the event horizon, though it is a Killing vector only at the horizon \cite{Chandrasekhar:1992}. The Komar conserved quantity at the event horizon associated with $\chi^{\mu}$ reads as \cite{Komar:1958wp}
\begin{eqnarray}
J_{\chi}&=&-\frac{1}{8\pi}\int_{S_t}\nabla^{\mu}\chi^{\nu}dS_{\mu\nu}, \nonumber\\
&=&-\frac{1}{8\pi}\int_{S_t}\nabla^{\mu}\left( \eta^{\mu}_{(t)}+\Omega \eta^{\mu}_{(\phi)}\right)dS_{\mu\nu}.
\end{eqnarray}
Using Eqs.~(\ref{mass1}) and (\ref{ang1}), we obtain
\begin{eqnarray}
J_{\chi}&=&M_{\text{eff}}-2\Omega J_{\text{eff}},\nonumber\\
&=&\frac{M(r_+^2-a^2)}{(r_+^2+a^2)}-\frac{\left(r_+^2-(s-1)a^2\right)}{(r_+^2+a^2)s}\frac{\Gamma}{ r_+^{-(s-2)/s}}.\label{ST}
\end{eqnarray}
To understand the implication of the above conserved quantity, we calculate the black hole horizon temperature \cite{Chandrasekhar:1992}
\begin{eqnarray}
T_+&=&\frac{\kappa}{2\pi}=\frac{\Delta'}{4\pi(r_+^2+a^2)},\nonumber\\
&=& \frac{(r_+-M)}{2\pi (r_+^2+a^2)}+\frac{( s-1)}{2\pi s(r_+^2+a^2)}\frac{\Gamma}{ r_+^{-(s-2)/s}},\label{temp}
\end{eqnarray}
whereas entropy is defined as follows:
\begin{equation}
S_+=\frac{A}{4}=\pi(r_+^2+a^2).\label{entropy}
\end{equation}
Equations (\ref{ST})$-$(\ref{entropy}) clearly infer that
\begin{equation}
J_{\chi}=M_{\text{eff}}-2\Omega J_{\text{eff}}=2S_+T_+.
\end{equation}
Therefore, the Komar conserved quantity corresponding to the null Killing vector at the event horizon $\chi^{\mu}$ is twice the product of the black hole entropy and the horizon temperature and hence satisfies the Smarr formula \cite{Smarr:1972kt,Bardeen:1973gs}.
\section{Black hole shadow}\label{sec4}
The light originating from either the luminous background or the accretion disk surrounding the black hole passes in the vicinity of the event horizon, and a part of it gets trapped inside the horizon while another part escapes to infinity. This results in the optical appearance of the black hole, namely the black hole shadow encircled by the bright photon ring \cite{Bardeen,Synge:1966,Luminet:1979,CT}. Synge \cite{Synge:1966}, in pioneering work, calculated the shadow cast by a Schwarzschild black hole, and thereafter Bardeen \cite{Bardeen} studied the shadow of Kerr black holes. In the past decade, shadows have been extensively studied for varieties of black holes \cite{De1,Grenzebach1}. Interestingly, it is found that the photon emission ring, i.e., the light rays that orbit around the black hole many times before they reach the distant observer, explicitly depends on the spacetime geometry, but is independent of the astrophysical details of the accretion flow model  \cite{Beckwith:2004ae,Johannsen:2010ru,Johannsen:2015qca}. Thus the structure of the photon ring encompassing the black hole shadow is a potential tool to test the signatures of strong gravitational lensing of nearby radiation, and, hence, its shape and size can reveal valuable information regarding the near-horizon field features of gravity. \\
For this purpose, we study the motion of a test particle in a stationary and axially symmetric black hole spacetime, which neglecting the back-reaction is completely defined by the four integrals of motion: the particle rest mass $m_0$, total energy $\cal E$, axial angular momentum $\cal L$, and the Carter constant $\mathcal{Q}$, related to the latitudinal motion of the test particle \cite{Carter:1968rr}. Using these integrals of motion, we obtained the null geodesics equation of motion in the first-order differential form \cite{Carter:1968rr,Chandrasekhar:1992}
\begin{eqnarray}
\Sigma \frac{dt}{d\tau}&=&\frac{r^2+a^2}{\Delta}\left({\cal E}(r^2+a^2)-a{\cal L}\right)  -a(a{\cal E}\sin^2\theta-{\mathcal {L}}) ,\label{tuch}\\
\Sigma \frac{dr}{d\tau}&=&\pm\sqrt{\mathcal{R}(r)} ,\label{r}\\
\Sigma \frac{d\theta}{d\tau}&=&\pm\sqrt{\Theta(\theta)} ,\label{th}\\
\Sigma \frac{d\phi}{d\tau}&=&\frac{a}{\Delta}\left({\cal E}(r^2+a^2)-a{\cal L}\right)-\left(a{\cal E}-\frac{{\cal L}}{\sin^2\theta}\right), \label{phiuch}
\end{eqnarray}
where $\tau$ is the affine parameter along the geodesics and 
\begin{eqnarray}\label{06}
\mathcal{R}(r)&=&\left((r^2+a^2){\cal E}-a{\cal L}\right)^2-\Delta ((a{\cal E}-{\cal L})^2+{\cal K}),\quad \\ 
\Theta(\theta)&=&{\cal K}-\left(\frac{{\cal L}^2}{\sin^2\theta}-a^2 {\cal E}^2\right)\cos^2\theta,\label{theta0}
\end{eqnarray}
and ${\cal K}$ stands for the separable constant which is related with Carter's constant of motion $\mathcal{Q}=\mathcal{K}+(a\mathcal{E}-\mathcal{L})^2$ \cite{Chandrasekhar:1992,Carter:1968rr}.
For $\mathcal{K}=0$, photon motions are restricted only to the equatorial plane. $\mathcal{R}(r)$ and $\Theta(\theta)$ are related to the effective potentials for the radial and the latitudinal motion of the photon, such that zeros of these potentials determine the turning point in the photon trajectories.
Let us define the dimensionless impact parameters
\begin{equation}
\eta\equiv{\cal K}/\mathcal{E}^2,\qquad \xi\equiv\mathcal{L}/\mathcal{E},
\end{equation} 
which characterize the null geodesics, such that, depending on their values, photons may undergo scattering orbits ($\eta>\eta_c$), capturing orbits ($\eta<\eta_c$), and unstable orbits ($\eta=\eta_c$), which are very crucial for the shadow formation and indeed mark the shadow silhouette. Thus, on the observer's celestial sky, the scattered photons account for the bright region, whereas captured photons attribute to the dark region. These unstable photon orbits, of constant radii $r_p$, witness continuum radial turning points, i.e., $\dot{r_p}=\ddot{r_p}=0$, corresponding to the extrema of effective potential
\begin{eqnarray}
\left.\mathcal{R}\right|_{(r=r_p)}=\left.\frac{\partial \mathcal{R}}{\partial r}\right|_{(r=r_p)}=0\;\; \text{and} \quad \left.\frac{\partial^2 \mathcal{R}}{\partial r^2}\right|_{(r=r_p)}> 0,
\end{eqnarray}
and form a photon region around the black hole. Furthermore, due to rotation of the black hole, photons can have either prograde motion or retrograde motion, whose respective radii $r_p^{-}$ and $r_p^{+}$ can be obtained as zeros of $\eta_c=0$. For the Kerr black hole, the photon orbit radii $r_p$ are
\begin{eqnarray}
r_p^-&=&2M\left[1+ \cos\left(\frac{2}{3}\cos^{-1}\left[-\frac{|a|}{M}\right]\right) \right],\nonumber\\
r_p^+&=&2M\left[1+ \cos\left(\frac{2}{3}\cos^{-1}\left[\frac{|a|}{M}\right]\right) \right],
\end{eqnarray}  
which for the Schwarzschild black hole ($a=0$), takes the degenerate value $r_p^-=r_p^+=3M$.
For the visualization of the black hole shadow, one has to consider the projection of the photon region onto the image plane. Thereby, the locus of the shadow boundary is defined in terms of two celestial coordinates $\alpha$ and $\beta$, which by construction lie in the celestial plane perpendicular to the line joining the observer and the center of the black hole and are related to the photon four-momentum $p^{(\mu)}$ measured in the orthonormal-tetrad basis \cite{Bardeen}. For an observer at position ($r_o,\theta_o$), in the far exterior region of the black hole, they read  \cite{CT}
\begin{equation}
\alpha=-r_o\frac{p^{(\phi)}}{p^{(t)}},\qquad \beta=r_o\frac{p^{(\theta)}}{p^{(t)}}.
\end{equation} 
On using geodesic Eqs.~(\ref{tuch}), (\ref{th}), and (\ref{phiuch}), the celestial coordinates yield
\begin{eqnarray}
\alpha&=&-\left. r_o\frac{\xi_c}{\sqrt{g_{\phi\phi}}(\zeta-\gamma\xi_c)}\right|_{(r_o,\theta_o)},\nonumber\\
\beta&=&\pm\left. r_o\frac{\sqrt{\Theta_{\theta}(\theta)}
}{\sqrt{g_{\theta\theta}}(\zeta-\gamma\xi_c)}\right|_{(r_o,\theta_o)},~\label{Celestial}
\end{eqnarray} 
with 
\begin{eqnarray}
\zeta=\sqrt{\frac{g_{\phi\phi}}{g_{t\phi}^2-g_{tt}g_{\phi\phi}}},\qquad \gamma=-\frac{g_{t\phi}}{g_{\phi\phi}}\zeta.
\end{eqnarray}
For an observer sitting in the asymptotically flat region ($r_o\to\infty$), the celestial coordinates Eq.~(\ref{Celestial}) can be simplified as \cite{Bardeen,CT}
\begin{equation}
\alpha=-\xi_c\csc\theta_o,\qquad \beta=\pm\sqrt{\eta_c+a^2\cos^2\theta_o-\xi_c^2\cot^2\theta_o}.\label{pt}
\end{equation} 
\begin{figure*}
	\begin{tabular}{c c}
		\includegraphics[scale=0.85]{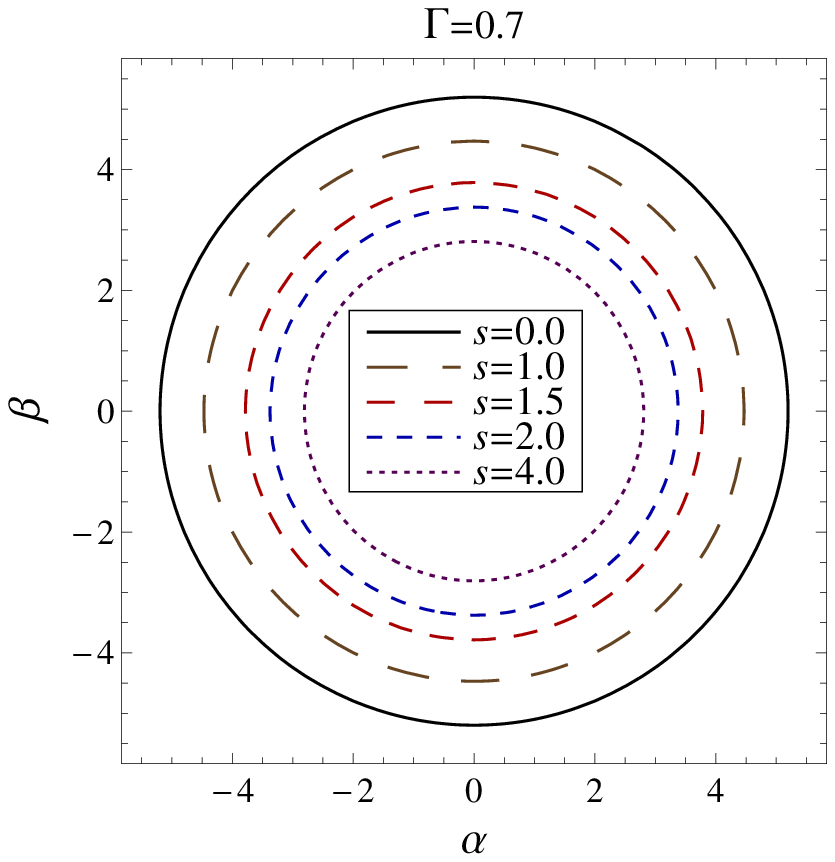}&
		\includegraphics[scale=0.85]{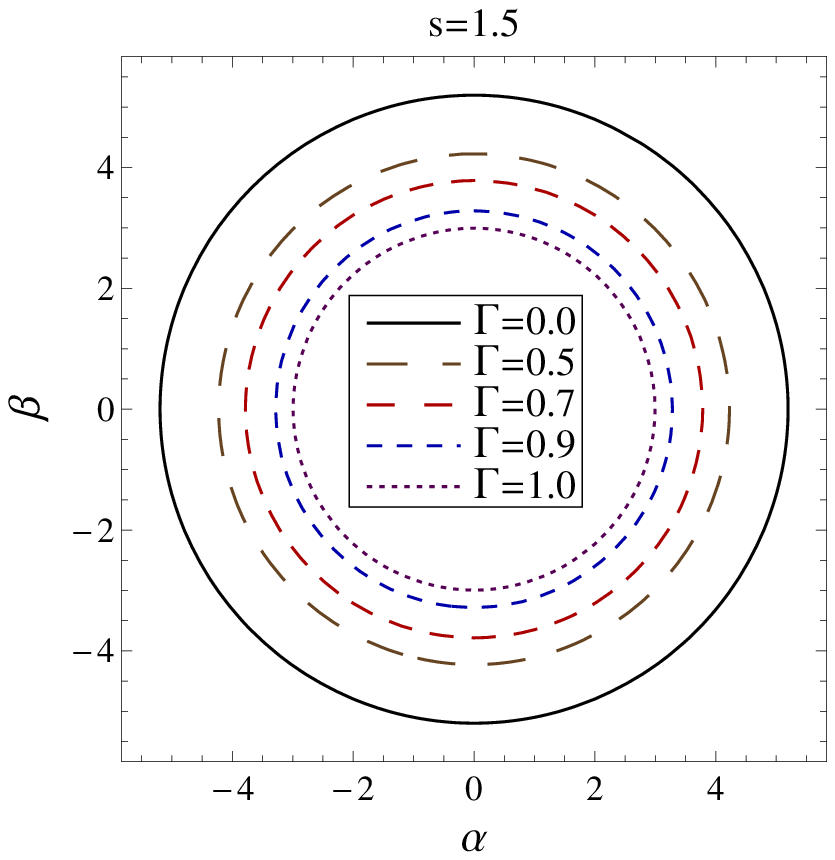}
	\end{tabular}
	\caption{Nonrotating Kalb-Ramond black hole shadows with varying parameters $s$ and $\Gamma$. The black solid line corresponds to the Schwarzschild black hole shadow. }
	\label{shadowNR}
\end{figure*}
We further consider that the observer is perceiving the black hole at an inclination angle $\theta_0=\pi/2$. Using geodesic Eqs. (\ref{r})-(\ref{phiuch}) and celestial coordinate (\ref{Celestial}), we obtain 
\begin{figure*}
	\begin{tabular}{c c}
		\includegraphics[scale=0.7]{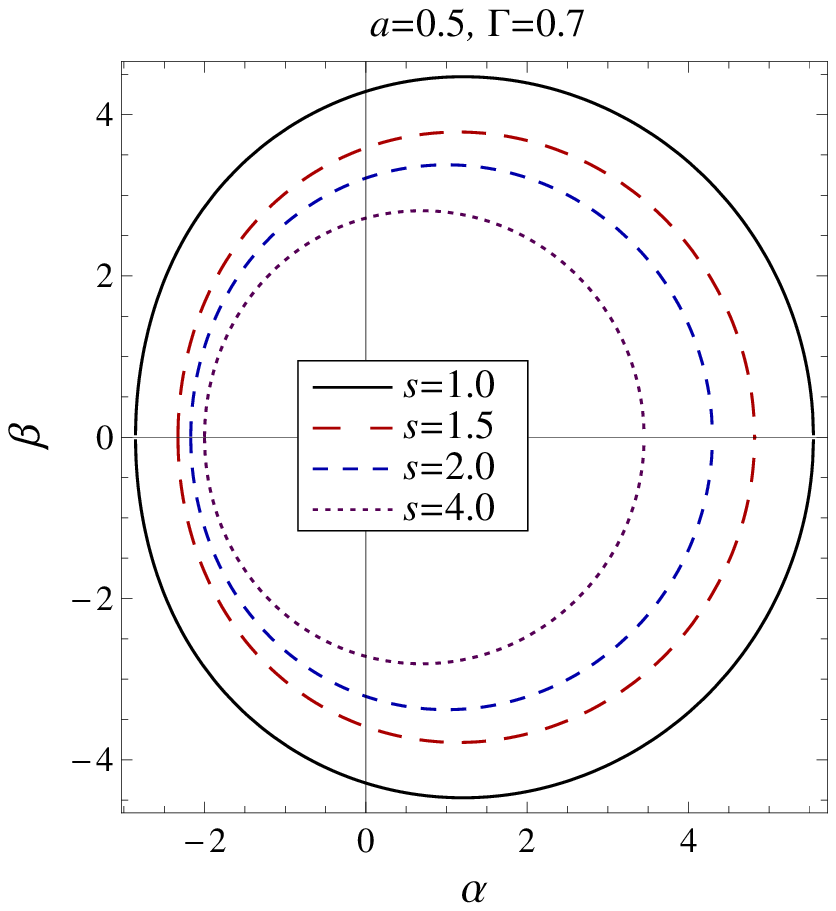}&
		\includegraphics[scale=0.7]{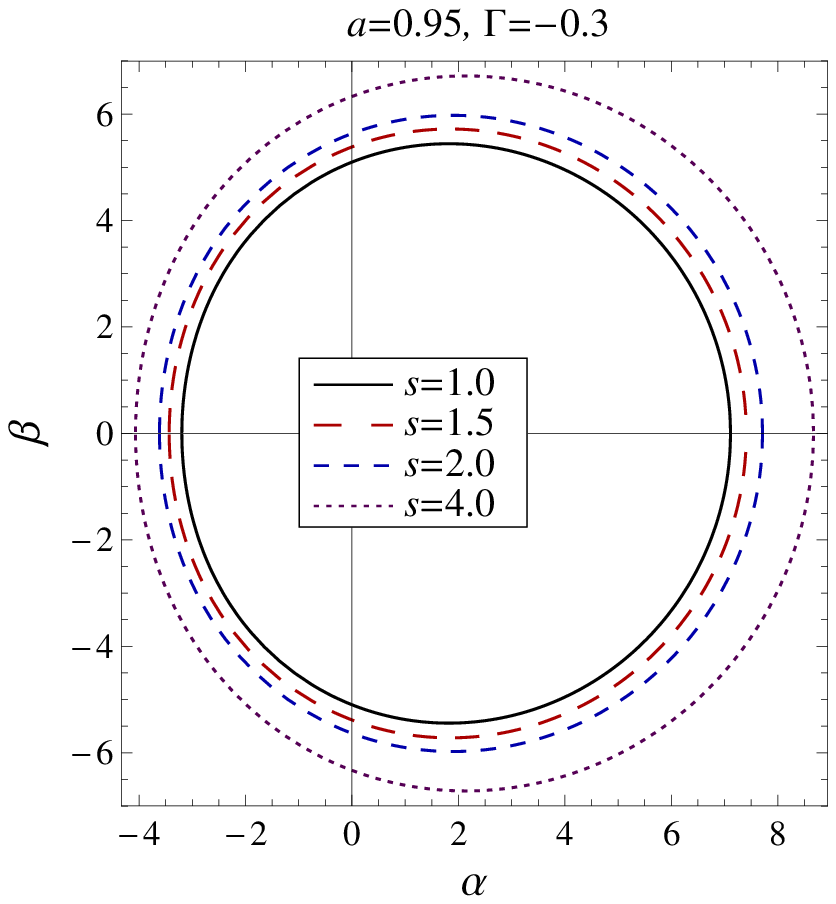}\\
		\includegraphics[scale=0.7]{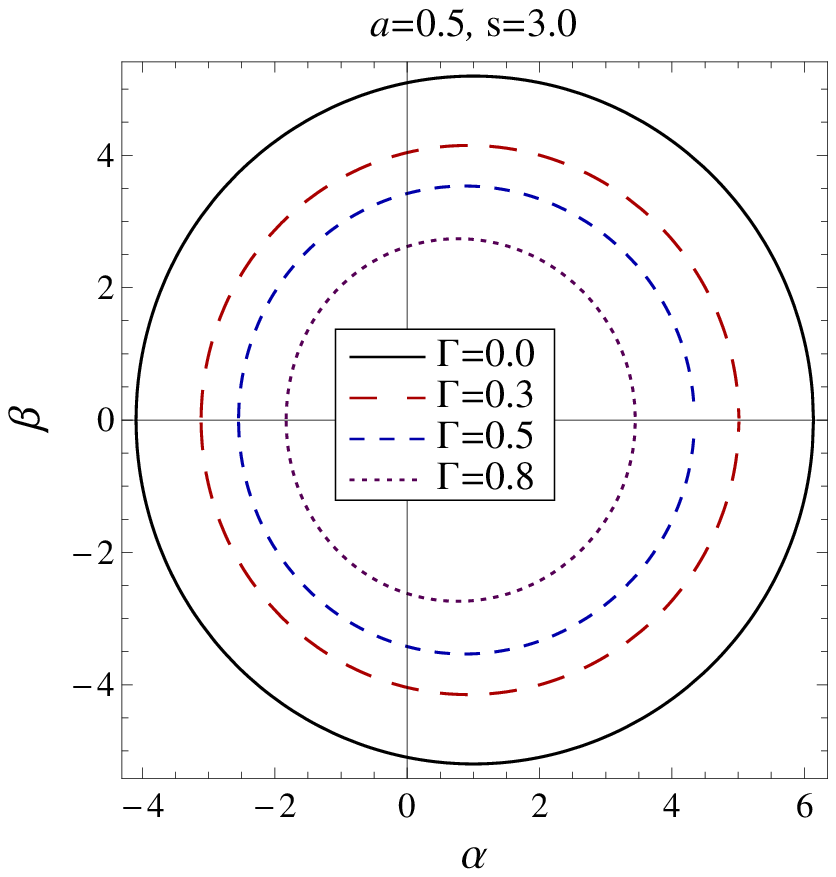}&
		\includegraphics[scale=0.7]{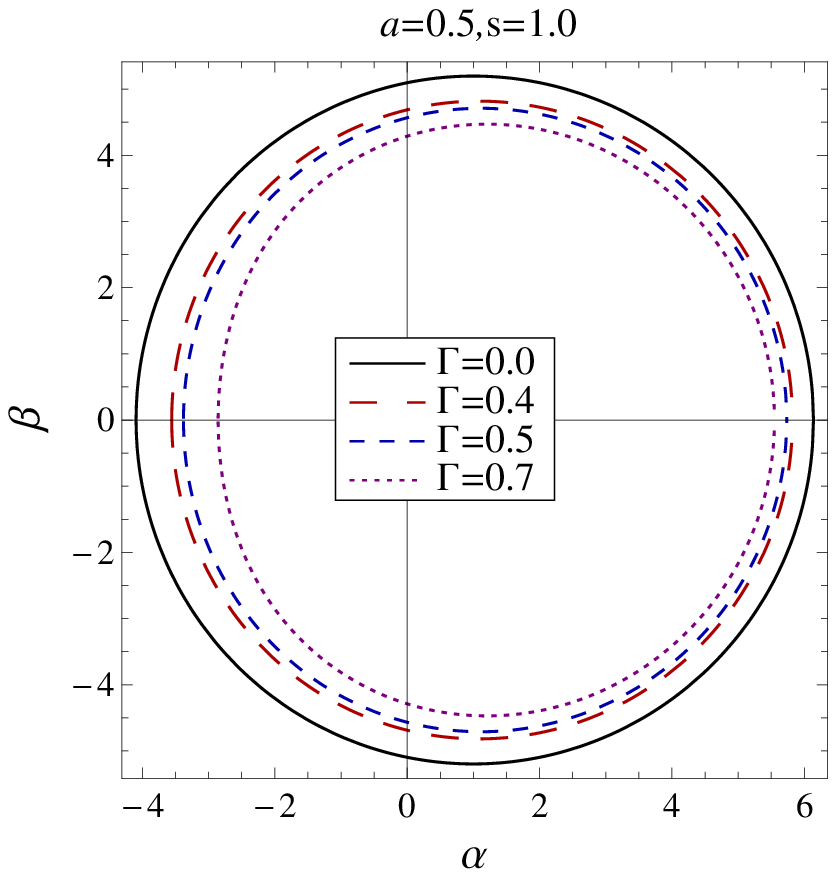}\\
		\includegraphics[scale=0.7]{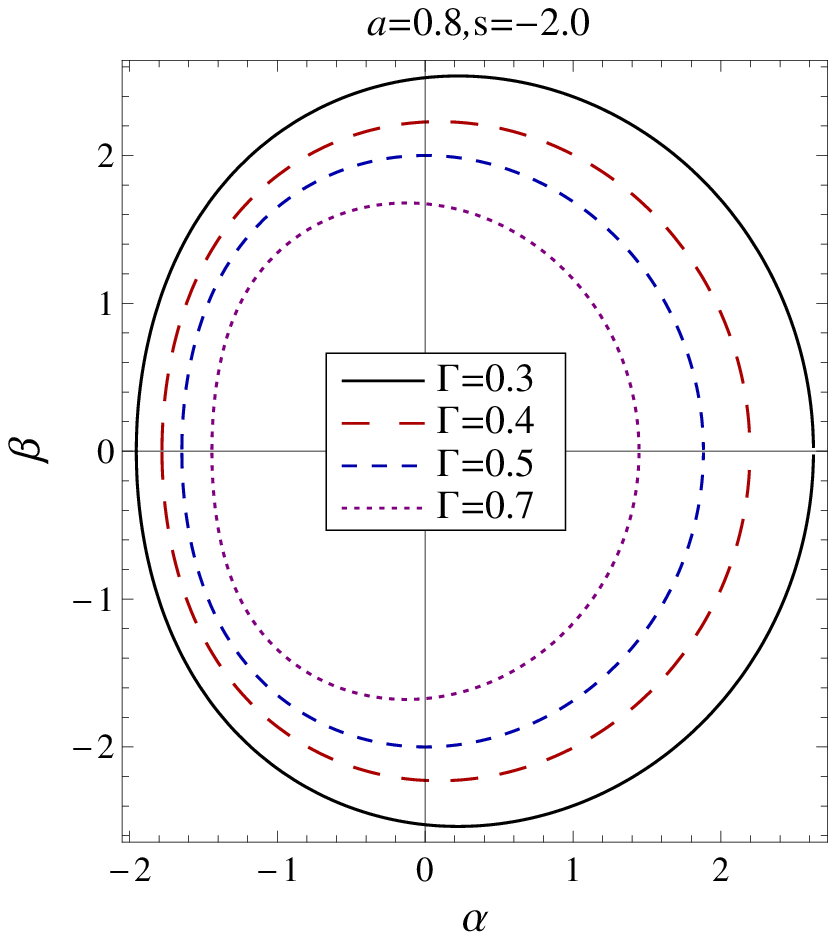}&
		\includegraphics[scale=0.7]{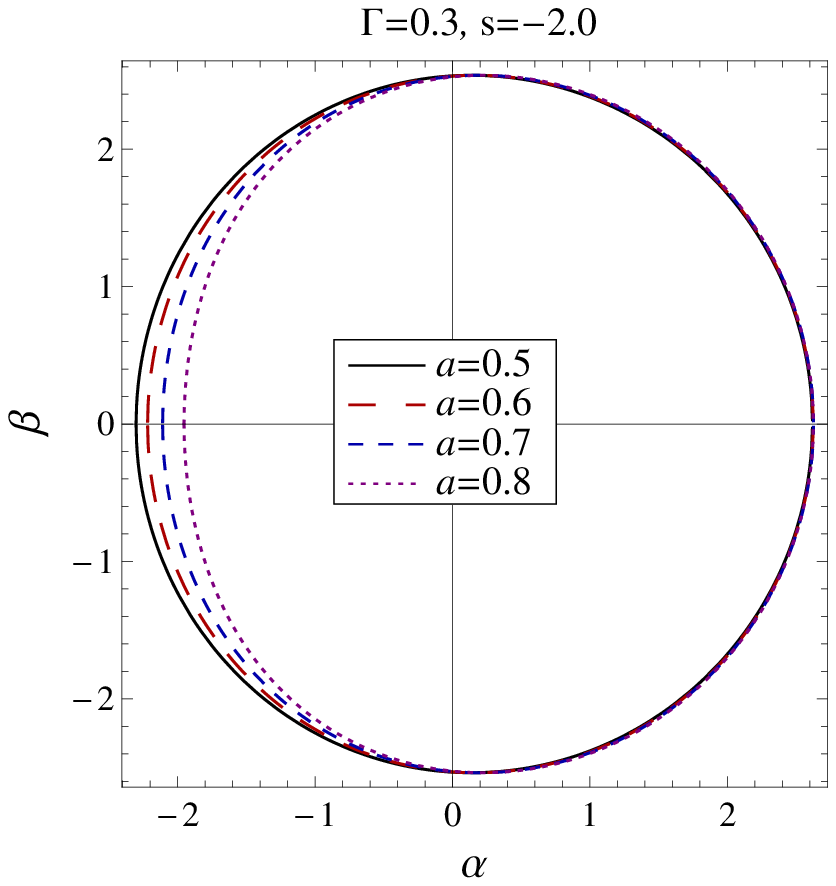}
	\end{tabular}
		\caption{Plot showing the rotating Kalb-Ramond black hole shadows with varying parameters $a$, $\Gamma$, and $s$. }
\label{shadow}
\end{figure*}
\begin{figure*}
	\begin{tabular}{c c c}
		\includegraphics[scale=0.63]{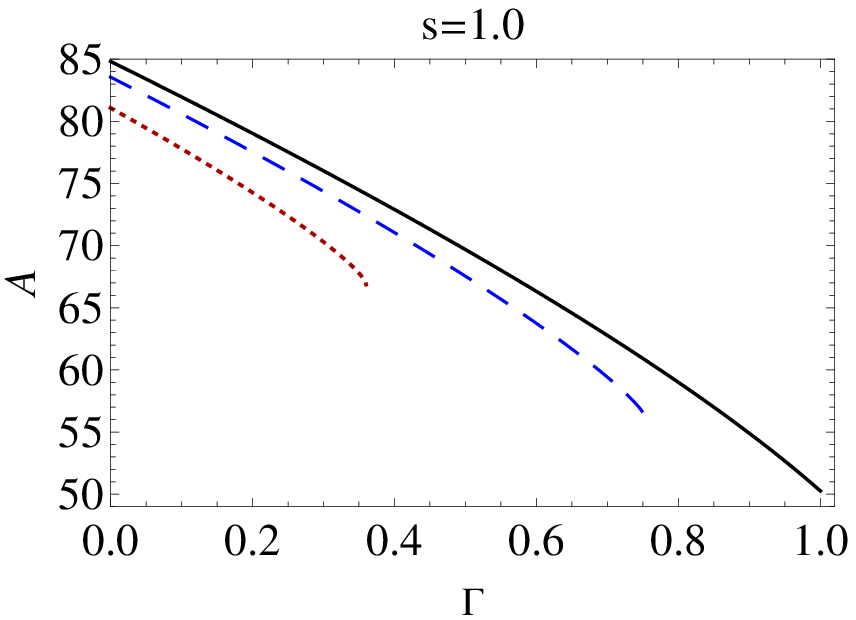}&
		\includegraphics[scale=0.63]{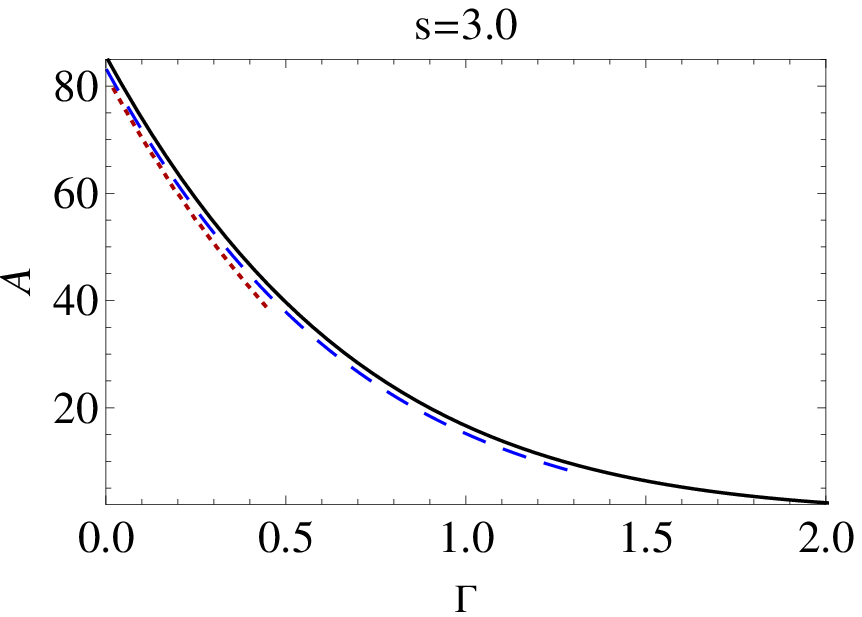}&
		\includegraphics[scale=0.63]{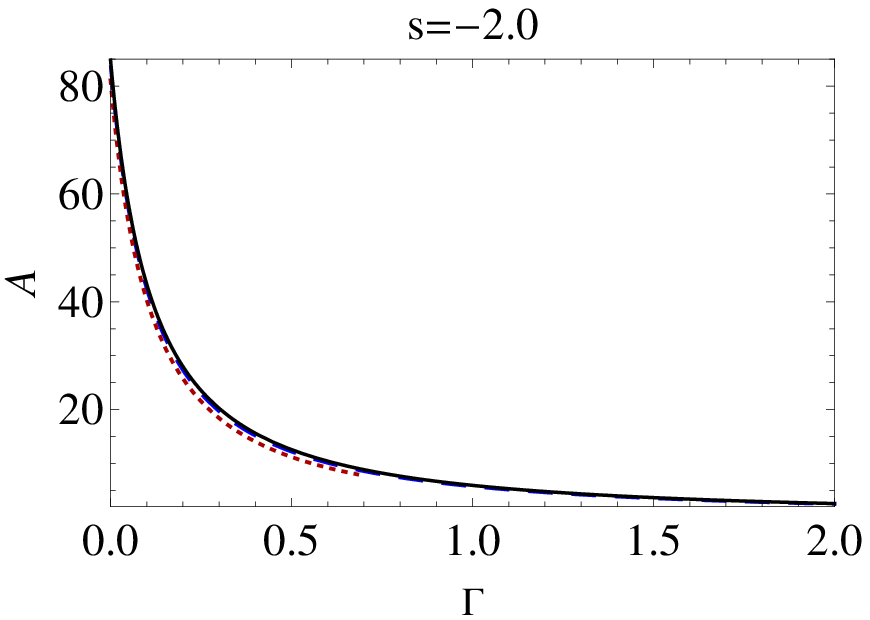}\\
		\includegraphics[scale=0.63]{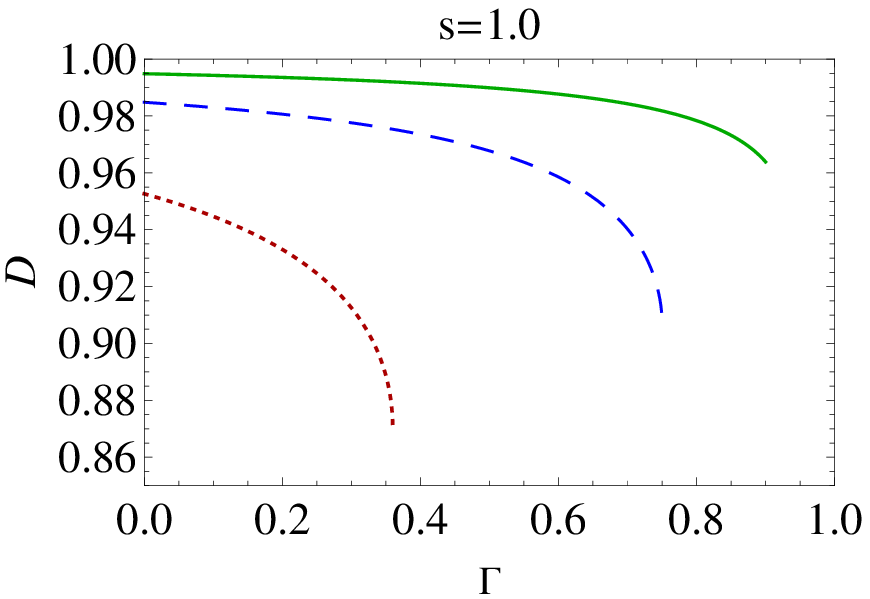}&
		\includegraphics[scale=0.63]{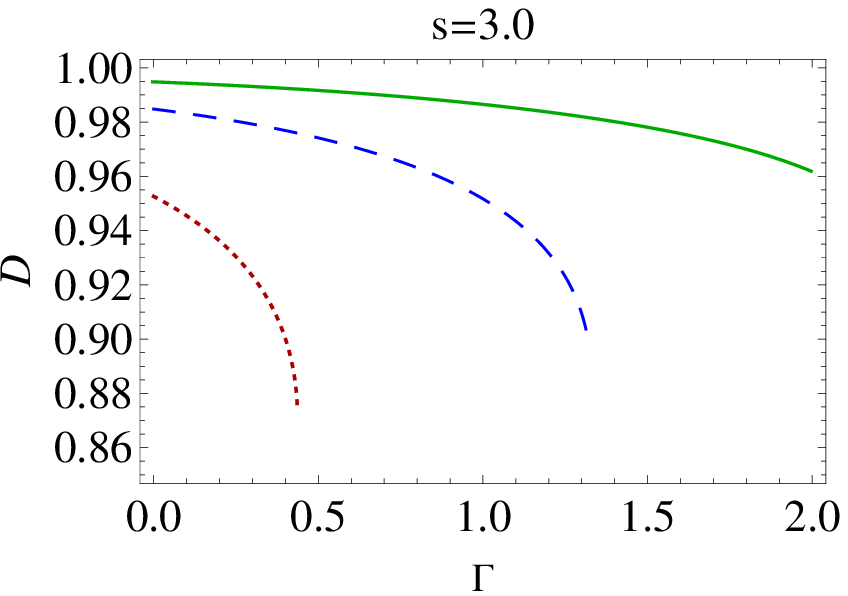}&
		\includegraphics[scale=0.63]{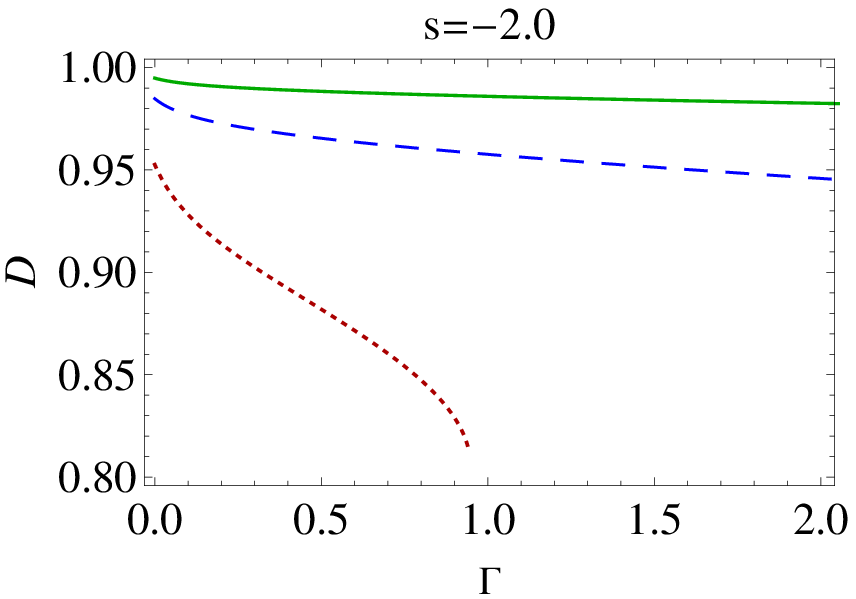}
	\end{tabular}
	\caption{Shadow area $A$ and oblateness observables $D$ vs $\Gamma$ for rotating Kalb-Ramond black holes (solid black curve) for the nonrotating black hole $a=0.0$, (solid green curve) for $a=0.3$, (dashed blue curve) for $a=0.5$, and (dotted red curve) for $a=0.8$.}
	\label{obs1}
\end{figure*}
\begin{figure*}
	\begin{tabular}{c c c}
		\includegraphics[scale=0.63]{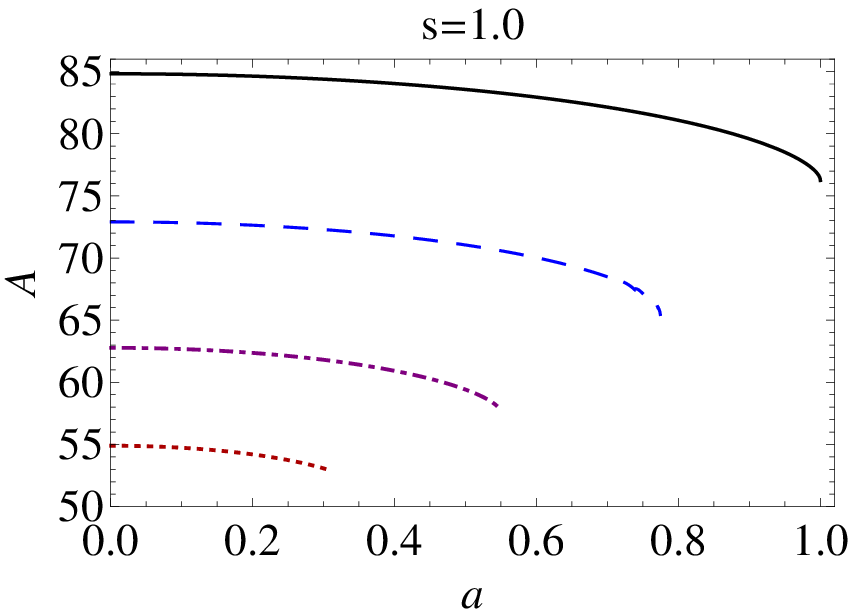}&
		\includegraphics[scale=0.63]{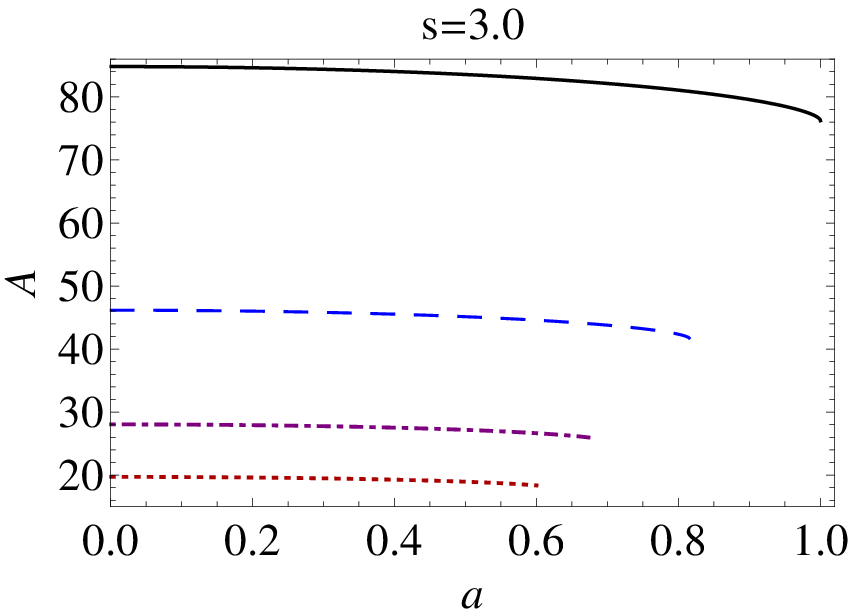}&
		\includegraphics[scale=0.63]{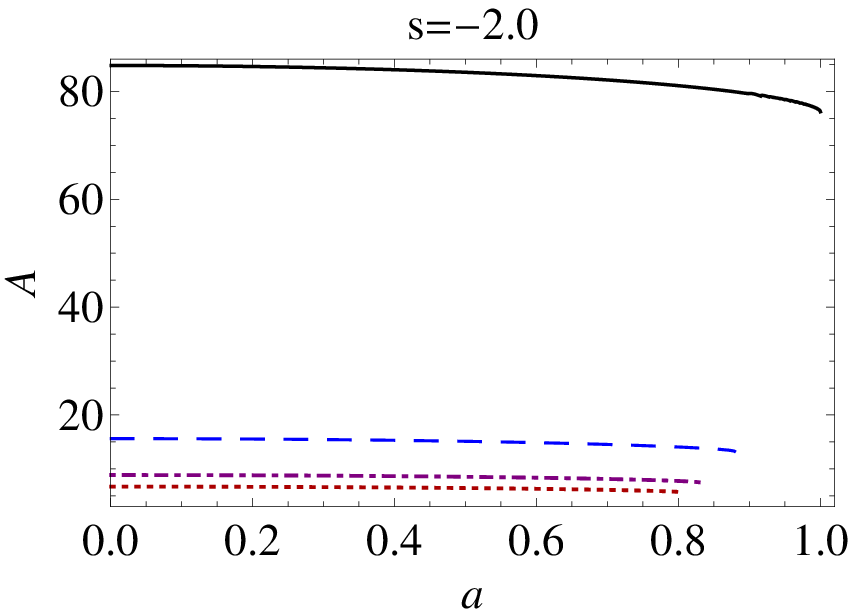}\\
		\includegraphics[scale=0.63]{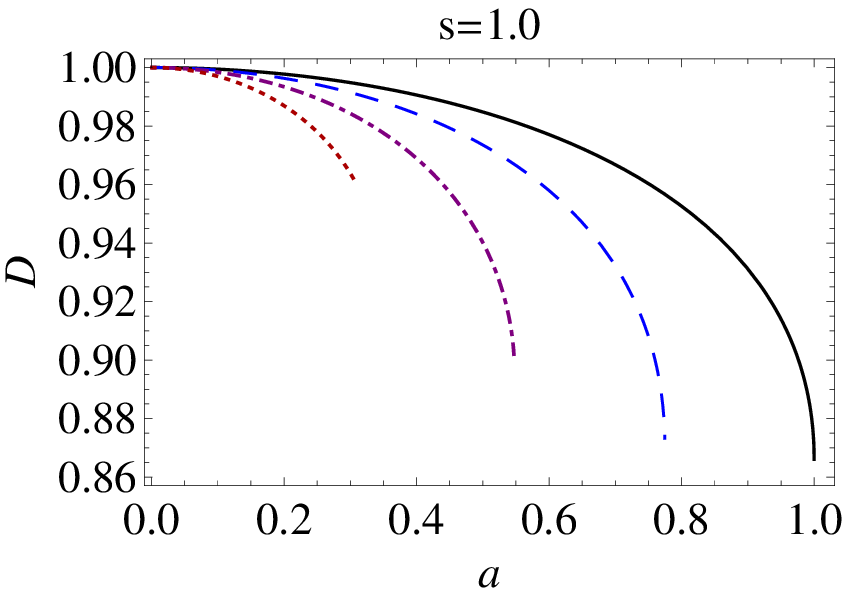}&
		\includegraphics[scale=0.63]{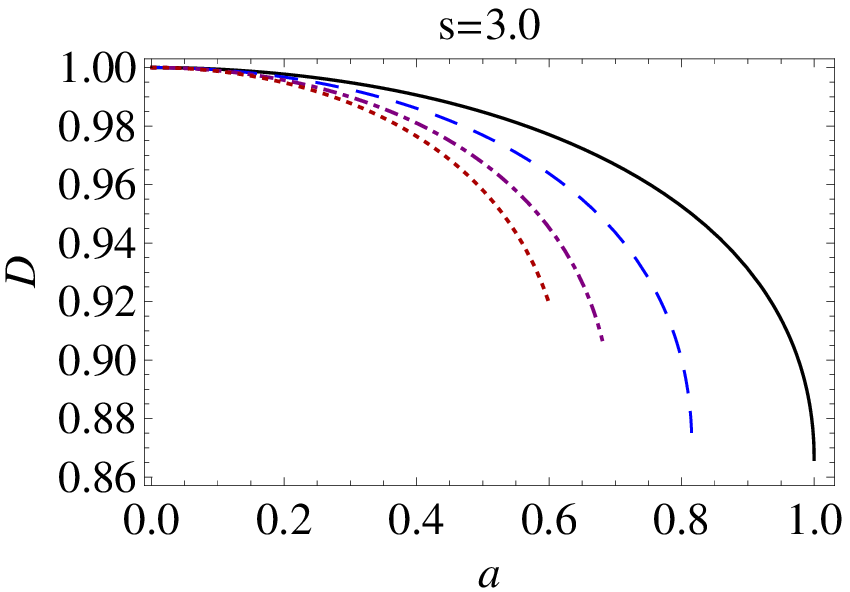}&
		\includegraphics[scale=0.63]{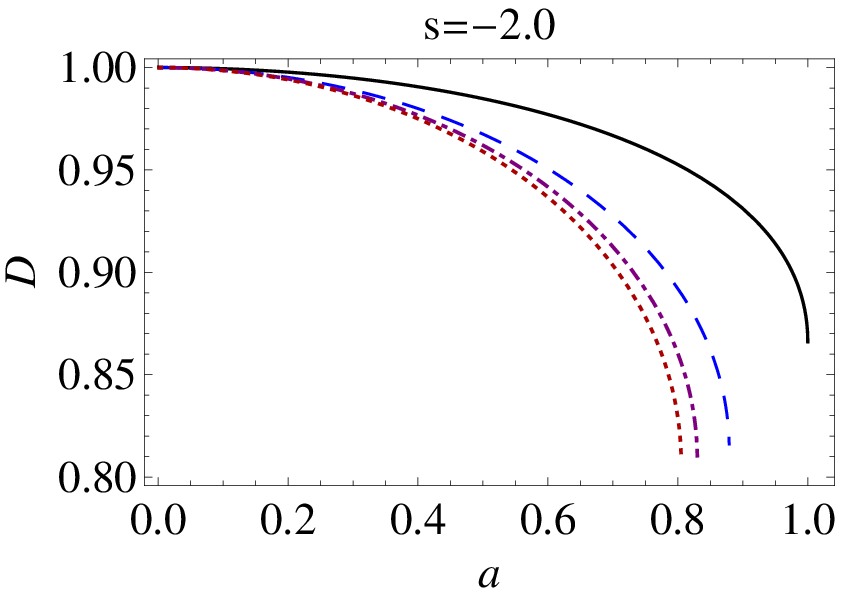}		
	\end{tabular}
	\caption{Shadow area $A$ and oblateness observables $D$ vs $a$ for rotating Kalb-Ramond black holes (solid black curve) for the Kerr black hole $\Gamma=0.0$, (dashed blue curve) for $\Gamma=0.4$, (dashdotted magenta curve) for $\Gamma=0.7$ and (dotted red curve) for $\Gamma=0.9$. }
	\label{obs2}
\end{figure*}

\begin{eqnarray}
\alpha&=& \frac{1}{a^2 [m(r_p) + r_p (-1 + m'(r_p))]^2}\Big[r_p^3 (-r_p^3 + m(r_p)\nonumber\\
&& (4 a^2 + 6 r_p^2 - 9 m(r_p) r_p ) \nonumber\\
&&- 2 r_p (2 a^2 + r_p^2 - 3 m(r_p) r_p) m'(r_p) - r_p^3 m'(r_p)^2)\Big],\nonumber\\
\beta&=&\frac{(a^2 - 3 r_p^2) m(r_p) + r_p (a^2 + r_p^2) (1 + m'(r_p))}{a (m(r_p)+ r_p (-1 + m'(r_p)))},\label{alpha}
\end{eqnarray}

\begin{figure*}
	\begin{tabular}{c c}
		\includegraphics[scale=0.78]{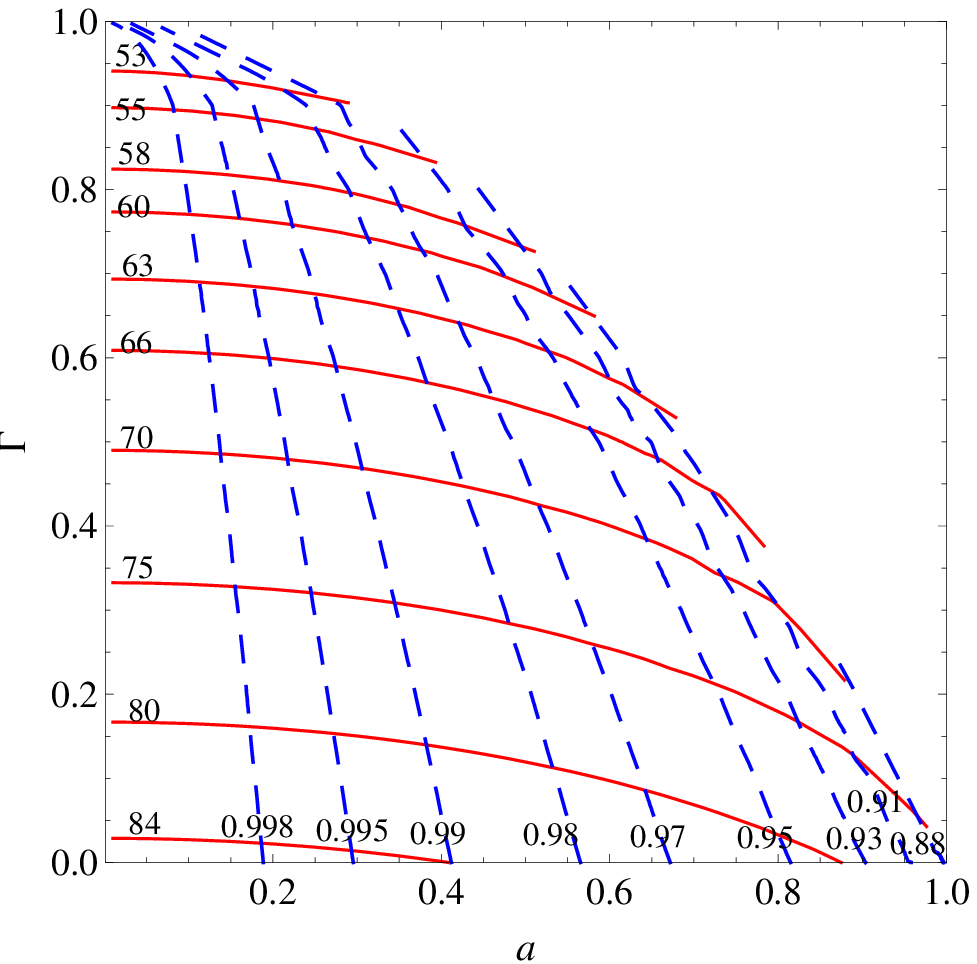}&
		\includegraphics[scale=0.78]{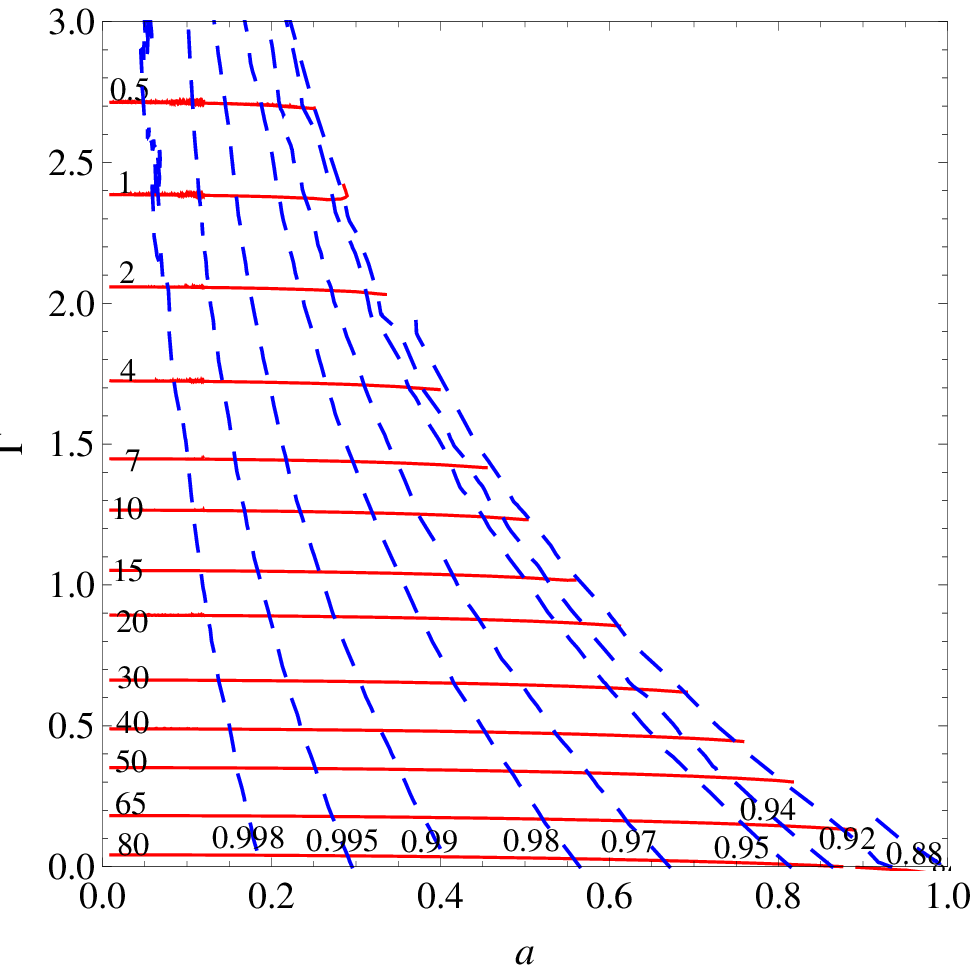}
	\end{tabular}
\caption{Contour plots of the observables $A$ and $D$ in the plane $(a, \Gamma)$ for the rotating Kalb-Ramond black hole (left panel) for $s=1$ and (right panel) for $s=3$. Each curve is labeled with the corresponding values of $A$ and  $D$. Solid red curves correspond to the area $A$, and dashed blue curves for oblateness $D$. }
	\label{obs3}
\end{figure*}
where, for brevity, we have defined
\begin{equation}
m(r)=M-\frac{\Gamma}{2r^{-(s-2)/s}}.\label{KRmass}
\end{equation}
For the nonrotating case, Eq.~(\ref{alpha}) yields 
\begin{equation}
\alpha^2+\beta^2=\frac{2r_p^2\left[4r_p^2-12M^2-3\frac{\Gamma^2}{r_p^{-2(s-2)/s}}+12M\frac{\Gamma}{r_p^{-(s-2)/s}}\right]}{(2M-\frac{\Gamma}{r_p^{-(s-2)/s}}-2r_p)^2},\label{nonrot}
\end{equation}
which clearly elucidates that for static spherically symmetric Kalb-Ramond black hole metric the shadow is circular in shape. For the Schwarzschild black hole ($a=0,s=0, r_p=3M$), Eq.~(\ref{alpha}) reduces to
\begin{equation}
\alpha^2+\beta^2=27M^2
\end{equation}
and infers that the shadow radius is $3\sqrt{3}M$.
Taking the unstable photon orbit radius $r_p$ as a parameter, the parametric plot $\beta$ vs $\alpha$ in Eq.~(\ref{alpha}) delineates the shadows for rotating Kalb-Ramond black holes.  Shadows of nonrotating Kalb-Ramond black holes are smaller than those for the Schwarzschild black holes, and the shadow size decreases with both increasing $s$ and $\Gamma$ (cf. Fig.~\ref{shadowNR}). Shadows of rotating Kalb-Ramond black holes for various values of black hole parameters are shown in Fig.~\ref{shadow}, which infers that the presence of the Kalb-Ramond field has a profound influence on the apparent shape and size of the shadow. For the characterization of shadows, we define two astronomical observables, namely shadow area $A$ and oblateness parameter $D$ \cite{Kumar:2018ple,Tsupko:2017rdo}
\begin{equation}
A=2\int{\beta(r_p) d\alpha(r_p)}=2\int_{r_p^{-}}^{r_p^+}\left( \beta(r_p) \frac{d\alpha(r_p)}{dr_p}\right)dr_p,\label{Area}
\end{equation} 
\begin{equation}
D=\frac{\alpha_r-\alpha_l}{\beta_t-\beta_b},\label{Oblateness}
\end{equation}
where $A$ and $D$, respectively, characterize the shadow size and shape. In Fig.~\ref{obs1}, shadow observables $A$ and $D$ are plotted with varying $\Gamma$ for different values of $s$ and $a$, and it is evident that shadow size decreases whereas distortion increases with increasing $\Gamma$ (cf. Fig. \ref{obs1}). The variation of $A$ and $D$ with spin parameter $a$ is shown in Fig.~\ref{obs2}. Moreover, the shadows of the rotating Kalb-Ramond black holes are smaller and more distorted than the corresponding Kerr black hole shadows ($\Gamma=0$ or $s=0$). The black solid curve in Fig. \ref{obs2} corresponds to the Kerr black hole. For the estimation of black hole parameters, we plotted these shadow observables in the ($a, \Gamma$) plane for different values of $s$ in Fig. \ref{obs3}. It is evident that each curve of constant $A$ and $D$ intersects at a unique point, which gives the value of black hole parameters $a$ and $\Gamma$.

The EHT Collaboration \cite{EHT} using the very large baseline interferometry technique has recently observed the central compact emission region at the center of galaxy M87 at the 1.3 mm wavelength, thereby opening a new window to test gravity in the strong-field regime \cite{Akiyama:2019cqa,Akiyama:2019eap,Akiyama:2019bqs,Akiyama:2019fyp}. The central flux depression by $\gtrsim 10:1$ and the asymmetric emission ring of crescent diameter $42\pm3 \mu$as in the captured image of the black hole M87* provide direct evidence of the black hole shadow, which is consistent with the predicted image of a Kerr black hole in general relativity \cite{Akiyama:2019cqa,Akiyama:2019eap,Akiyama:2019bqs,Akiyama:2019fyp}. The observed shadow of the M87* black hole has been used to constrain or rule our various black hole models in general relativity as well in modified gravities \cite{Bambi1}. We can use the relevant shadow observable, the asymmetry parameter $\Delta C$, to constrain the parameter space of rotating Kalb-Ramond black holes. The shadow boundary can be described by a one-dimensional closed curve characterized by the radial and angular coordinates ($R(\varphi),\varphi$) in a polar coordinate system with the origin at the shadow center ($\alpha_C,\beta_C$). The shadow average radius $\bar{R}$ is defined by \cite{Johannsen:2010ru}
\begin{equation}
\bar{R}=\frac{1}{2\pi}\int_{0}^{2\pi} R(\varphi) d\varphi,
\end{equation}
with
\begin{equation}R(\varphi)= \sqrt{(\alpha-\alpha_C)^2+(\beta-\beta_C)^2},\;\ \varphi\equiv \tan^{-1}\left(\frac{\beta}{\alpha-\alpha_C}\right)\nonumber.
\end{equation}
\begin{figure*}
	\begin{tabular}{c c}
		\includegraphics[scale=0.73]{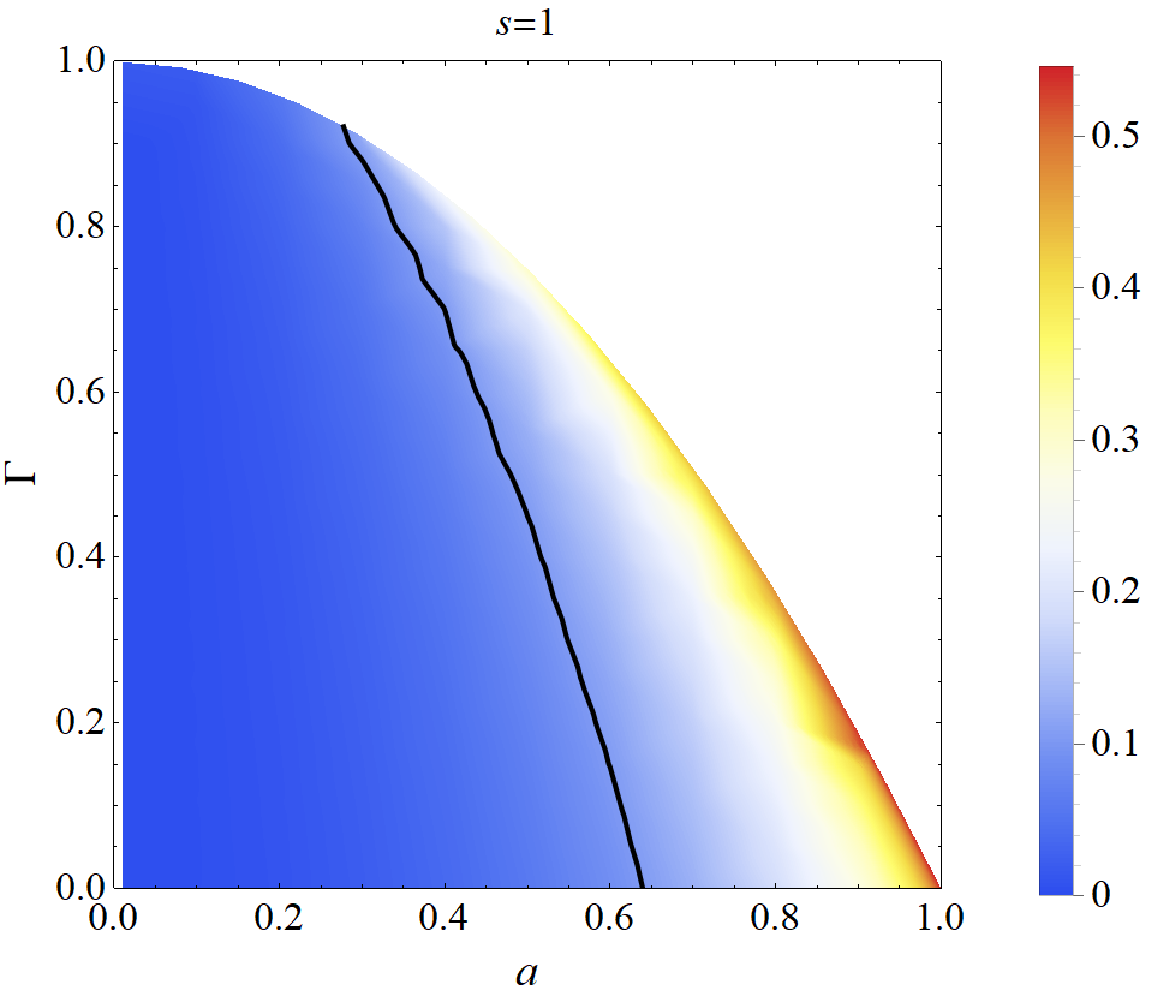}&
		\includegraphics[scale=0.73]{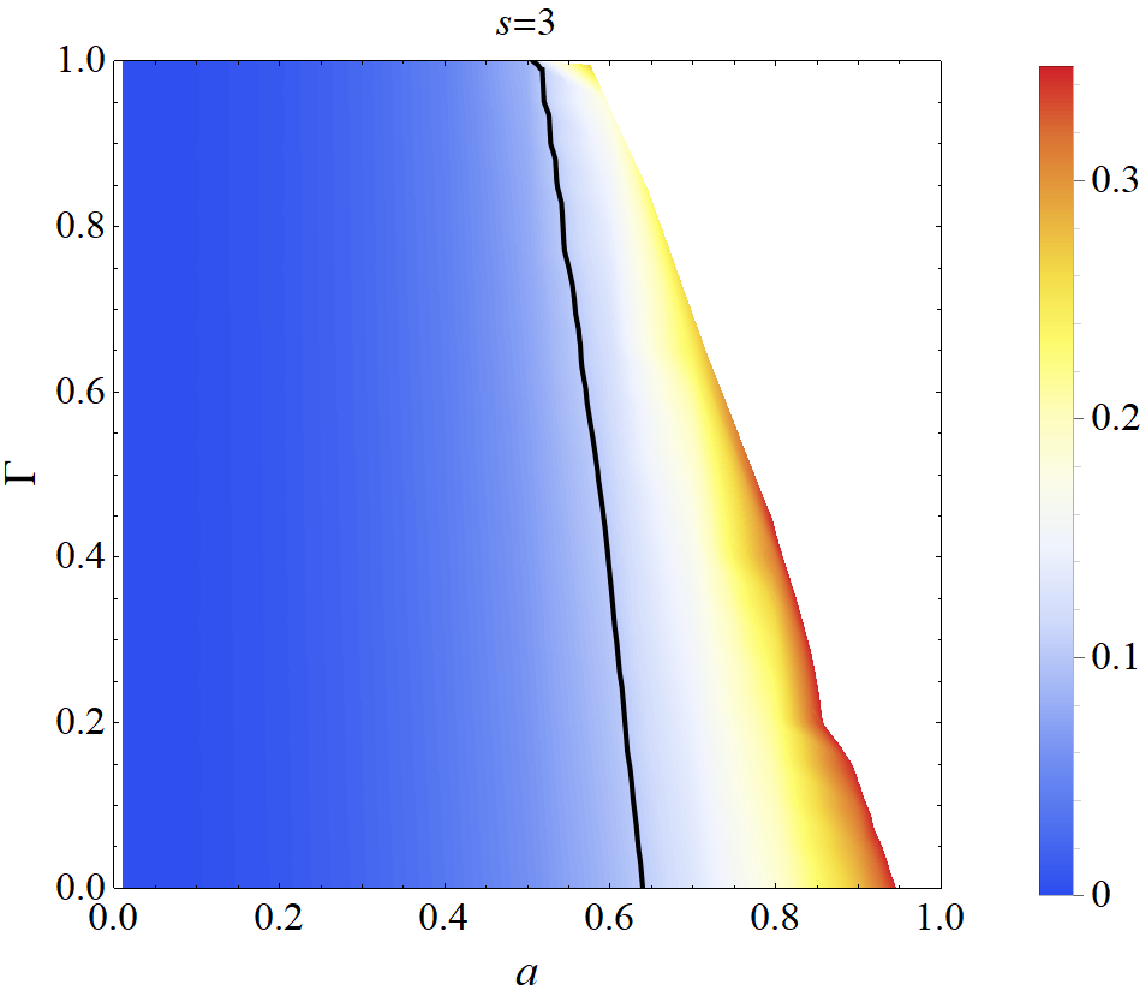}
	\end{tabular}
	\caption{Deviation from circularity $\Delta C$ for rotating Kalb-Ramond black holes shadows as a function of parameters ($a,\Gamma$). The black solid line corresponds to $\Delta C=0.10$, such that the region above the black line is excluded by the measured circularity of the M87* black hole reported by the EHT, $\Delta C\leq 0.10$.  }
	\label{M87}
\end{figure*}
\begin{figure*}
	\begin{tabular}{c c}
		\includegraphics[scale=0.73]{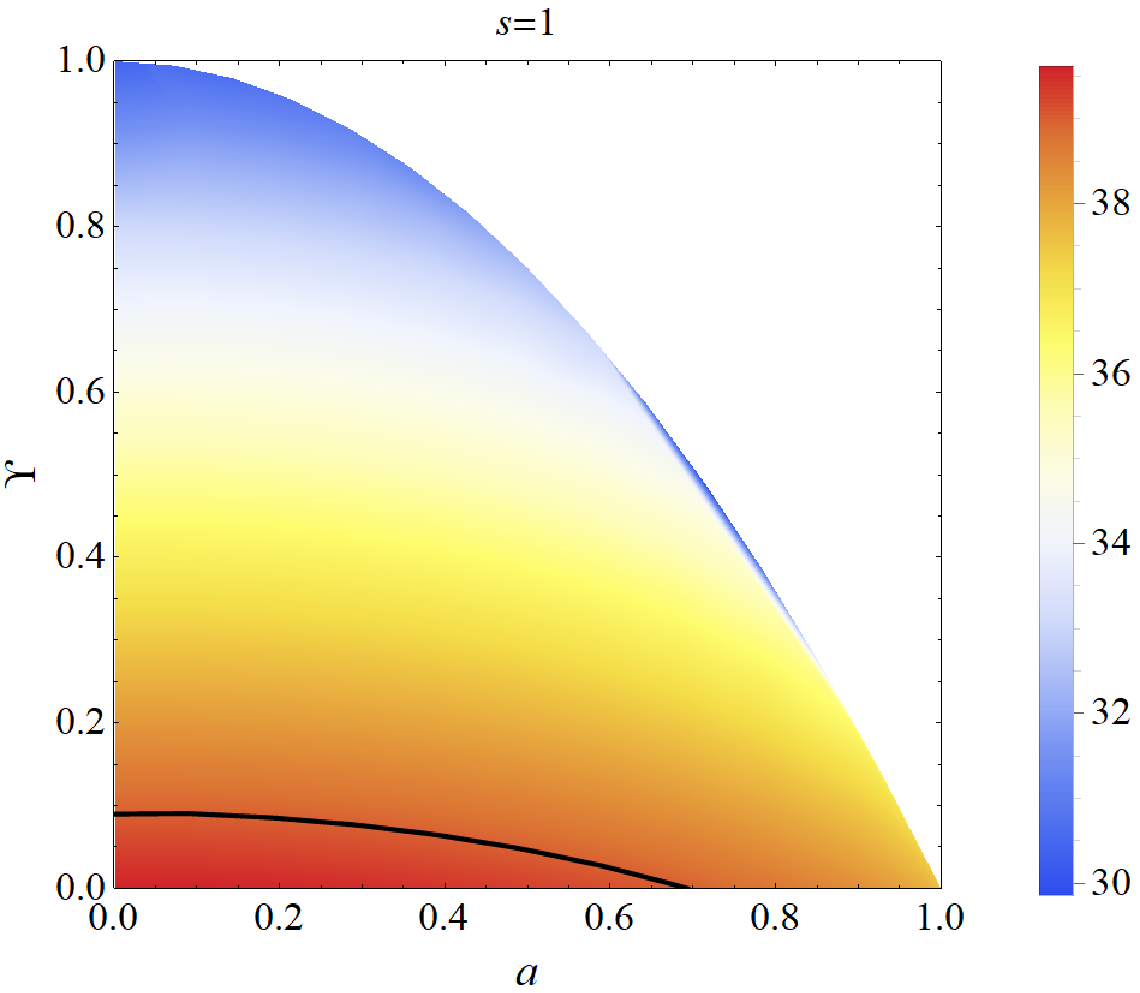}&
		\includegraphics[scale=0.73]{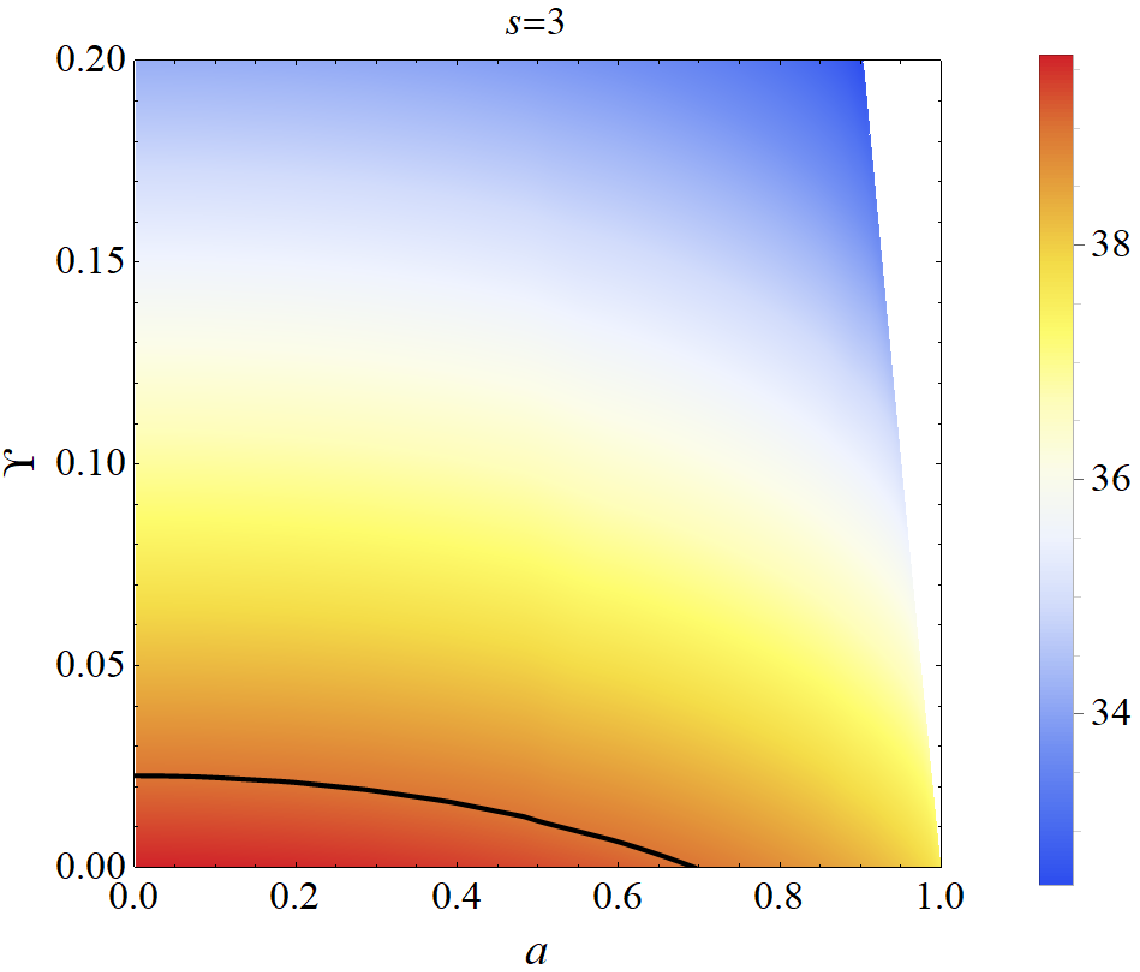}
	\end{tabular}
	\caption{Shadow angular diameter $\theta_d$ for rotating Kalb-Ramond black holes as a function of parameters ($a,\Gamma$). The black solid line corresponds to $\theta_d=39\,\mu$as, such that the region below the black line falls within the $1\sigma$ region of the measured angular diameter of the M87* black hole reported by the EHT,  $\theta_d=42\pm 3\,\mu$as.   }
	\label{M87A}
\end{figure*}
The circularity deviation $\Delta C$ measures the deviation from a perfect circle and defined in terms of the root-mean-square distance from the average radius \cite{Johannsen:2010ru,Johannsen:2015qca}
\begin{equation}
\Delta C=2\sqrt{\frac{1}{2\pi}\int_0^{2\pi}\left(R(\varphi)-\bar{R}\right)^2d\varphi},
\end{equation}
such that, for a perfect circular shadow, $\Delta C$ identically vanishes. Tracing the emission ring, the EHT deduced that the circularity deviation in the observed image of the M87* black hole is $\Delta C\leq 0.10$ \cite{Akiyama:2019cqa}. We calculate the circularity deviation for metric Eq.~(\ref{rotbhtr}) and use the EHT bound to put constraints on the black hole and the Kalb-Ramond field parameters. The interplay between black hole spin $a$ and field parameters $\Gamma$ and $s$ for the shadow asymmetry parameter $\Delta C$ is shown in Fig~\ref{M87}. It is evident that $\Delta C$ merely constrained the $\Gamma$. 

The black hole shadow with areal radius $R_s=\sqrt{A/\pi}$ deduces an angular diameter $\theta_d$ on the observer's celestial sky:
\begin{equation}
\theta_d=2\frac{R_s}{d},
\end{equation}    
where $d$ is the black hole distance from Earth, $d=16.8$ Mpc for M87*. The emission ring diameter in the observed M87* black hole shadow is $\theta_d=42\pm 3\, \mu$as. The angular diameter for the rotating Kalb-Ramond black hole shadow is calculated and shown as a function of  $\Gamma$ and $a$ in Fig.~\ref{M87A} for $s=3$ and $s=1$, considering $M=6.5\times 10^9 M_{\odot}$. The region enclosed by the black solid line, $\theta_d=39\,\mu$as, falls within the $1\sigma$ region of the M87* shadow angular diameter. Figure \ref{M87A} infers that the M87* shadow angular size sufficiently constrains the black hole parameters ($a,\Gamma$); however, the constraints on the $\Gamma$ are more strong for large values of $s$, $s=3$. 

\section{Gravitational deflection of light}\label{sec5}
\begin{figure}
	\includegraphics[scale=0.52]{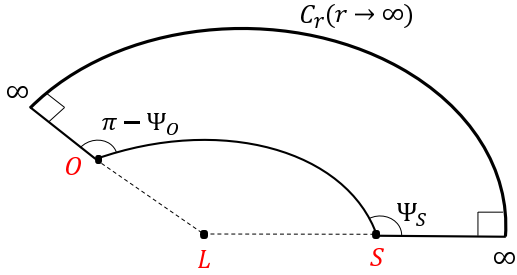}
\caption{Schematic figure for the quadrilateral ${}_O^{\infty}\Box_{S}^{\infty}$ embedded in the curved space. Light emitted by the source $S$ gets deflected by the black hole $L$ and reaches the observer $O$. }\label{lensing1}
\end{figure}

Gibbon and Werner \cite{Gibbons:2008rj} used the Gauss-Bonnet theorem, which connects the differential geometry of the surface with its topology, in the context of optical geometry to calculate the deflection angle of light in a spherically symmetric black hole spacetime \cite{Carmo}. Later, Ishihara \textit{et al.} \cite{Ishihara:2016vdc}, taking into account the finite distance from the black hole to a light source and an observer, calculate the light deflection angle in static, spherically symmetric and asymptotically flat spacetimes, which is generalized by Ono, Ishihara, and Asada \cite{Ono:2017pie} for stationary and axisymmetric spacetimes, and, later, extensively used for varieties of black hole spacetimes \cite{Crisnejo1:2018uyn}. We follow their approach to calculate the light deflection angle in the weak-field limit caused by the rotating Kalb-Ramond black hole. We assume that both the observer ($O$) and the source ($S$) are at a finite distance from the black hole ($L$) (cf. Fig.~\ref{lensing1}). The deflection angle at the equatorial plane can be defined in terms of the angle made by light rays at the source and observer $\Psi_S$ and $\Psi_O$, respectively, and their angular coordinate separation $\Phi_{OS}$ \cite{Ono:2017pie}:
\begin{equation}
\alpha_D=\Psi_O-\Psi_S+\Phi_{OS}.
\end{equation}
Here, $\Phi_{OS}=\Phi_O-\Phi_S$, where $\Phi_O$ and $\Phi_S$ are, respectively, the angular coordinates of the observer and the source. We consider a three-dimensional Riemannian manifold $^{(3)}\mathcal{M}$ defined by optical metric $\gamma_{ij}$, in which photon motion is described as a spatial curve \cite{Gibbons:2008rj}. To calculate the deflection angle using the Gauss-Bonnet theorem, we consider a quadrilateral ${}_O^{\infty}\Box_{S}^{\infty}$, the domain of integration, embedded in the curved space $^{(3)}\mathcal{M}$ which consists of a spatial light ray curve from the source to the observer, a circular arc segment $C_r$ of coordinate radius $r_C$ $(r_C\to\infty)$, and two
outgoing radial lines from $O$ and from $S$ (cf. Fig.~\ref{lensing1}). The Gauss-Bonnet theorem yields the geometrically invariant definition as follows \cite{Ono:2017pie}:
\begin{equation}
\alpha_D=-\int\int_{{}_O^{\infty}\Box_{S}^{\infty}} K dS+\int_{S}^{O} k_g dl,\label{deflectionangle}
\end{equation}
where $K$ is the Gaussian curvature of the two-dimensional surface of light propagation and $k_g$ is the geodesic curvature of light curves, a measure for the deviation of curve from the geodesics. $dS$ and $dl$ are, respectively, the infinitesimal area element of the surface and arc length  element. Since Eq.~(\ref{deflectionangle}) is invariant in differential geometry, $\alpha_D$ is well defined even if focal point $L$ is a singularity \cite{Ishihara:2016vdc}. For the null geodesics $ds^2=0$, we get
\begin{equation}
dt= \pm\sqrt{\gamma_{ij}dx^i dx^j}+N_i dx^i,
\end{equation}
with
\begin{eqnarray}
\gamma_{ij}dx^i dx^j&=&\frac{\Sigma^2}{\Delta(\Delta-a^2\sin^2\theta)}dr^2+\frac{\Sigma^2}{\Delta-a^2\sin^2\theta}d\theta^2\nonumber\\ 
&+& \left(r^2+a^2+\frac{2m(r)ra^2\sin^2\theta}{\Delta-a^2\sin^2\theta}\right)\frac{\Sigma\sin^2\theta\, d\phi^2}{(\Delta-a^2\sin^2\theta)} ,\nonumber\\
N_idx^i&=&-\frac{2m(r)ar\sin^2\theta}{\Delta-a^2\sin^2\theta}d\phi.\label{metric3}
\end{eqnarray}
An optical (or spatial) metric defined in this way gives the arc length ($l=\gamma_{ij}dx^{i}dx^{j}$), where $l$ is the affine parameter along the light curve \cite{Asada:2000vn}.
The deflection angle $\alpha_D$ defined in Eq.~(\ref{deflectionangle}) has a contribution from the curvature of the surface of light propagation $^{(3)}\mathcal{M}$ and the geodesics curvature of light curves as well. The Gaussian curvature of the surface is defined as \cite{Werner:2012rc}
\begin{align}
K=&\frac{{}^{3}R_{r\phi r\phi}}{\gamma}\nonumber\\
=&\frac{1}{\sqrt{\gamma}}\left(\frac{\partial}{\partial \phi}\left(\frac{\sqrt{\gamma}}{\gamma_{rr}}{}^{(3)}\Gamma^{\phi}_{rr}\right) - \frac{\partial}{\partial r}\left(\frac{\sqrt{\gamma}}{\gamma_{rr}}{}^{(3)}\Gamma^{\phi}_{r\phi}\right)\right),
\end{align}
where $\gamma=\det(\gamma_{ij})$. For a generic rotating and axially symmetric metric Eq.~(\ref{rotbhtr}), the Gaussian curvature $K$ is computed as 
 \begin{eqnarray}
K&=&\frac{-1}{6r^5(r-2m(r))}\Big(6r^2\left(r-2m(r)\right)(\Delta+a^2)m''(r)\nonumber\\
&&+6rm'(r)(rm'(r)-m(r))(\Delta+5a^2)+6rm(r)^2\nonumber\\
&&-(7r^2+a^2)m(r)+2r(r^2+3a^2) \Big),\label{K}
\end{eqnarray}
and, using the Kalb-Ramond black hole mass function defined in Eq.~(\ref{KRmass}) and considering a special case of $s=1$,  Eq.~(\ref{K}) yields
\begin{eqnarray}
K&=&\frac{3\Gamma}{r^4}+\frac{2\Gamma^2}{r^6}+\frac{8\Gamma a^2}{r^6}-\frac{3\Gamma^3}{2 r^8}-\frac{6\Gamma^2 a^2}{r^8}-\Big(\frac{2}{r^3}+\frac{6\Gamma}{r^5}\nonumber\\
&&+\frac{6a^2}{r^5}-\frac{12\Gamma a^2}{r^7}\Big)M+\Big(\frac{3}{r^4}-\frac{6 a^2}{r^6}+\frac{30 \Gamma a^2}{r^8}\Big)M^2\nonumber\\
&&-\frac{12M^3a^2}{r^7} -\left(\frac{24a^2}{r^8}-\frac{112\Gamma}{r^8}\right)M^4\nonumber\\
&&+\mathcal{O}\left(\frac{M\Gamma^2a^2}{r^9},\frac{M^3\Gamma a^2}{r^9}\right).
\end{eqnarray}
A fully consistent analytic treatment of the metric Eq.~(\ref{rotbhtr}) will involve an expansion in the powers of ($1/r$), which will lead to a very complicated expression. We exclusively work in the weak-field limit, ensuring that it captures
all the effects of the Kalb-Ramond field, and consider only the leading-order contributing terms.
The surface integral of Gaussian curvature over the closed quadrilateral ${}_O^{\infty}\Box_{S}^{\infty}$ reads \cite{Ono:2017pie}
\begin{equation}
\int\int_{{}_O^{\infty}\Box_{S}^{\infty}} K dS= \int_{\phi_S}^{\phi_O}\int_{\infty}^{r_0} K \sqrt{\gamma}dr d\phi,\label{Gaussian}
\end{equation}
where $r_0$ is the distance of closest approach to the black hole. The boundary of integration domain, namely the curve from $S$ to $O$ in the quadrilateral ${}_O^{\infty}\Box_{S}^{\infty}$, is unknown \textit{a priori}; hence, we first obtain the light orbit equation using Eqs.~(\ref{r}) and (\ref{phiuch}), that  reads 
\begin{equation}
\left(\frac{du}{d\phi}\right)^2=F(u),\label{orbit}
\end{equation}
with 
\begin{eqnarray} F(u)&=&\frac{(1+a^2u^2-(2M-\Gamma u)u)^2}{[(2M-\Gamma u)u(a-b)+b]^2} \Big[1+(a^2-b^2)u^2\nonumber\\
&&+(2M-\Gamma u)u^3(a-b)^2\Big],\label{fu}
\end{eqnarray}
where $u=1/r$ and $b\equiv \xi$ is the impact parameter. In the weak-field approximation, we obtain the solution  \cite{Crisnejo:2019ril}
\begin{eqnarray}
u&=&\frac{\sin\phi}{b} + \frac{M(1-\cos\phi)^2}{b^2}-\frac{2Ma(1-\cos\phi)}{b^3}\nonumber\\
&-&\frac{M^2(60\phi\,\cos\phi+3\sin3\phi-5\sin\phi)}{16b^3}\nonumber\\
&+&\frac{a^2\sin^3\phi}{2b^3} +\mathcal{O}\left( \frac{M\Gamma}{b^4},\frac{M^3}{b^4}\right)
,\label{uorbit}
\end{eqnarray} 
and the integral Eq.~(\ref{Gaussian}) can be recast as
\begin{equation}
\int\int_{{}_O^{\infty}\Box_{S}^{\infty}} K dS= \int_{\phi_S}^{\phi_O}\int_{0}^{u}-\frac{K\sqrt{\gamma}}{u^2}du d\phi,
\end{equation}
which for the rotating metric Eq.~(\ref{metric3}) reads as

\begin{widetext}
\begin{eqnarray}
\int\int K dS&=&\left(\cos^{-1}bu_o+\cos^{-1}bu_s\right)\Big(-\frac{3\Gamma}{4b^2} -\frac{21\Gamma a^2}{16b^4}+\frac{15\Gamma^2}{64b^4} +\frac{15M^2}{4b^2}-\frac{4M^2a}{b^3}+\frac{105M^2a^2}{16b^4}-\frac{15M^2\Gamma }{b^4}\Big)\nonumber\\
&+&\left(\sqrt{1-b^2u_o^2}+\sqrt{1-b^2u_s^2}\right)\Big(\frac{2M}{b}+\frac{2Ma^2}{b^3}-\frac{13 M\Gamma }{3b^3}+\frac{6M\Gamma a}{b^4}\Big) \nonumber\\
&+& \left(u_o\sqrt{1-b^2u_o^2}+u_s\sqrt{1-b^2u_s^2} \right)\Big(-\frac{3\Gamma}{4b} -\frac{21\Gamma a^2}{16b^3}+\frac{15\Gamma^2}{64b^3}-\frac{M^2}{4b}+\frac{81 M^2a^2 }{16b^3}-\frac{39 M^2\Gamma }{16b^3}-\frac{3 M^2\Gamma a}{b^4}\Big)\nonumber\\
&+& \left(u_o^2\sqrt{1-b^2u_o^2} +u_s^2\sqrt{1-b^2u_s^2}  \right)\Big(-\frac{ 2M\Gamma}{3b}+\frac{ Ma^2}{b}+\frac{43M\Gamma^2 }{30b^3}-\frac{191 M\Gamma a^2}{30b^3}-\frac{5 M\Gamma^2 a}{3b^4}\Big)\nonumber\\
&+& \left(u_o^3\sqrt{1-b^2u_o^2} +u_s^3\sqrt{1-b^2u_s^2}  \right)\Big(-\frac{7\Gamma a^2}{8b}+\frac{5\Gamma^2}{32b}+\frac{5\Gamma^2 a^2}{12b^3}-\frac{15M^2\Gamma}{16b}+\frac{15M^2a^2}{8b}+\frac{325M^2\Gamma^2}{64b^3}-\frac{45M^2\Gamma^2a}{8b^4}\Big)\nonumber\\
&+&\mathcal{O}\left(\frac{M^3}{b^3},\frac{M^2a\Gamma}{b^5}\right).\label{Gaussian1}
\end{eqnarray}
\end{widetext}

Here, $u_o$ and $u_s$, respectively, are the inverse of the $r_o$ and $r_s$, the distances of observer and source from the black hole, and we have used $\cos\phi_o=-\sqrt{1-b^2u_o^2},\;\; \cos\phi_s=\sqrt{1-b^2u_s^2}$; the negative sign is because the source and the observer are at the opposite sides to the black hole. Interestingly, a term linear in $a$ appears in Eq.~(\ref{Gaussian1}); thus, the contribution from the Gaussian curvature is sensitive to the direction of black hole rotation. Next, we calculate the geodesic curvature of the light curves, which is the surface-tangential component of the acceleration of the parameterized curve and given by
\begin{equation}
k_g=-\frac{1}{\sqrt{\gamma\gamma^{\theta\theta}}}N_{\phi,r}.
\end{equation}
For the metric (\ref{metric3}), this reads
\begin{eqnarray}
k_g&=&-\frac{2Ma}{r^3}+\frac{2a\Gamma}{r^4}-\frac{2M^2a}{r^4}-\frac{3M^3a}{r^5}+\frac{3Ma\Gamma}{r^5}\nonumber\\
&&-\frac{a\Gamma^2}{r^6}+\frac{6M^2a\Gamma}{r^6}-\frac{5M^4a}{r^6}-\frac{15Ma\Gamma^2}{4r^7}+\frac{25M^3a\Gamma}{2r^7}\nonumber\\
&&+\frac{152M^5a}{r^7}+\mathcal{O}\left(\frac{M^2a\Gamma^2}{r^8},\frac{M^4a\Gamma}{r^8},\frac{M^6a}{r^8}\right)
\end{eqnarray}
which identically vanishes for the nonrotating black hole. For the path integral of the $k_g$ along the light curve in Eq.~(\ref{deflectionangle}), the line element $dl$ is given by
\begin{equation}
dl=\sqrt{\left( \gamma_{rr}\left(\frac{dr}{d\phi}\right)^2+\gamma_{\phi\phi}\right)}d\phi.
\end{equation}
It is worthwhile to note that the geodesic curvatures of the curves from $S$ to $S_{\infty}$ and from $O$ to $O_{\infty}$ in Fig.~\ref{lensing1} are both zero, since these paths are geodesics, whereas $k_g$ is the geodesic curvature of the photon rays from $S$ to $O$, which is a spatial curve. Using Eqs.~(\ref{metric3}), (\ref{fu}), and (\ref{uorbit}), this gives
\begin{widetext} 
\begin{eqnarray}
\int_S^O k_g dl &=& \left(\cos^{-1}bu_o+\cos^{-1}bu_s\right)\Big(\frac{a\Gamma}{2b^3} -\frac{6M^2a}{b^3} +\frac{6M^2a^2}{b^4} \Big)+\left(\sqrt{1-b^2u_o^2}+\sqrt{1-b^2u_s^2}\right)\Big(-\frac{2Ma}{b^2}+\frac{6Ma\Gamma}{b^4}\Big)\nonumber\\
&+&\left(u_o\sqrt{1-b^2u_o^2}+u_s\sqrt{1-b^2u_s^2} \right)\Big(\frac{a\Gamma}{2b^2}-\frac{9 a\Gamma^2 }{16b^4}-\frac{2M^2a}{b^2}+\frac{2M^2a^2}{b^3}+\frac{115M^2a\Gamma}{8b^4}\Big)\nonumber\\
&+& \left(u_o^2\sqrt{1-b^2u_o^2} +u_s^2\sqrt{1-b^2u_s^2}  \right)\Big(\frac{ 2Ma\Gamma}{b^2}-\frac{4Ma^2\Gamma}{3b^3}-\frac{2Ma\Gamma^2}{b^4}\Big)\nonumber\\
&+&\left(u_o^3\sqrt{1-b^2u_o^2} +u_s^3\sqrt{1-b^2u_s^2}  \right)\Big(-\frac{3a\Gamma^2}{8b^2}+\frac{45M^2a\Gamma }{8b^2}-\frac{15M^2a^2\Gamma}{2b^3}\Big)+\mathcal{O}\left(\frac{M^3}{b^3}, \frac{M^2a\Gamma}{b^5}\right).\label{geodesiccurvature}
\end{eqnarray}   
\end{widetext}
Here, we have assumed that $dl>0$ such that the orbital angular momentum of the photons is aligned along the black hole spin; for other cases, $dl<0$ can be taken which will lead to an extra ``-" sign in Eq.~(\ref{geodesiccurvature}). 
Using Eqs.~(\ref{Gaussian1}) and (\ref{geodesiccurvature}) in Eq.~(\ref{deflectionangle}), we obtain the analytical expression for the gravitational deflection angle of light in the rotating Kalb-Ramond black hole spacetime Eq.~(\ref{rotbhtr}), which leads to a very lengthy expression. Nevertheless, in the asymptotic limits $u_o\to 0$ and $u_s\to 0$, i.e., the source and observer are at a very far distance from the black hole, the deflection angle for the rotating Kalb-Ramond black hole takes a rather simpler form as follows:
\begin{eqnarray}
\alpha_D&=&\left.\alpha_D\right|_{\text{Kerr}}-\frac{3\pi\Gamma}{4b^2}+\frac{a\pi\Gamma}{2b^3}-\frac{32M\Gamma}{3b^3}+\frac{76Ma\Gamma}{3b^4}\nonumber\\
&&-\frac{15\pi M^2\Gamma}{b^4}-\frac{21\pi a^2\Gamma}{16b^4}+\frac{15\pi \Gamma^2}{64b^4}\nonumber\\
&&+\mathcal{O}\Big(\frac{M\Gamma^2}{b^5},\frac{M^2a\Gamma}{b^5},\frac{M^3\Gamma}{b^5}\Big),~\label{deflection}
\end{eqnarray}
where $\left.\alpha_D\right|_{\text{Kerr}}$ corresponds to the Kerr deflection angle and reads as
\begin{eqnarray}
\left.\alpha_D\right|_{\text{Kerr}}&=&\frac{4M}{b} -\frac{4Ma}{b^2}+ \frac{15\pi M^2}{4b^2}-\frac{10\pi M^2a}{b^3}+\frac{4Ma^2}{b^3}\nonumber\\
&&+\frac{128M^3}{3b^3}+\mathcal{O}\left(\frac{M^4}{b^4},\frac{M^3a}{b^4}\right),
\end{eqnarray}
which is in agreement with the recent results on the deflection of light in the Kerr background with the higher-order contributing terms \cite{Crisnejo:2019ril,Iyer:2009wa,Edery:2006hm}.
The deflection angle for the nonrotating Kalb-Ramond black hole can be obtained as a special case of $a=0$ from Eq.~(\ref{deflection}):
\begin{align}
\alpha_D=&\frac{4M}{b}+\frac{15\pi M^2}{4b^2}-\frac{3\pi\Gamma}{4b^2}-\frac{32M\Gamma}{3b^3}+\frac{128 M^3}{3b^3}\nonumber\\
&-\frac{15\pi M^2\Gamma}{b^4}+\frac{15\pi \Gamma^2}{64b^4}+\mathcal{O}\left(\frac{M^4}{b^4},\frac{M\Gamma^2}{b^5},\frac{M^5}{b^5} \right),
\end{align}
which further reverts the value for the Schwarzschild black hole in the limiting case of $\Gamma=0$ \cite{Virbhadra:1999nm,Crisnejo:2019ril} as
\begin{equation}
\left.\alpha_D\right|_{\text{Schw}}=\frac{4M}{b}+\frac{15\pi M^2}{4b^2}+\frac{128M^3}{3b^3}+\mathcal{O}\left(\frac{M^4}{b^4} \right).
\end{equation}
Setting up the premises for the gravitational lensing, next we discuss the possible astronomical implications of the rotating black holes in the presence of the Kalb-Ramond field background. 
\begin{table}
	\begin{tabular}{|c|c|c|c|c|c|}
		\hline
		$\Gamma $  & $a=0.1$    & $a=0.3$     & $a=0.5$     &$ a=0.7 $   &$ a=0.9 $\\  
		\hline\hline
		0.1&    0.0486227& 0.0486161& 0.0486096& 0.048603& 0.0485965 \\  \hline
		0.3&    0.145868& 0.145848& 0.145829& 0.145809& 0.145789 \\  \hline
		0.5&    0.243113& 0.243081& 0.243048& 0.243015& 0.242982 \\  \hline
		0.7&    0.340359& 0.340313& 0.340267& 0.340221& 0.340175 \\  \hline
		0.9&   	0.437604& 0.437545& 0.437486& 0.437427& 0.437368 \\  \hline
		1.1&   	0.534849& 0.534777& 0.534705& 0.534633& 0.534561 \\  \hline
	\end{tabular}
	\caption{The corrections in the deflection angle $\delta\alpha_D =\left.\alpha_D\right|_{\text{Kerr}}-\alpha_D$ for Sgr A* with $b=10^3M$, source at $r_s=10^5M$, and varying $\Gamma$ and $a$; $\delta\alpha_D$ is in units of arcsec.}\label{T1}
\end{table}
\begin{table}[h!]
	\begin{tabular}{|c|c|c|c|c|c|}
		\hline
		$\Gamma $  & $a=0.1$    & $a=0.3$     & $a=0.5$     &$ a=0.7 $   &$ a=0.9 $   \\
		\hline	\hline
		0.1&    0.0440628& 0.0440569& 0.044051& 0.044045& 0.0440391 \\  \hline
		0.3&    0.132188& 0.132171& 0.132153& 0.132135& 0.132117 \\  \hline
		0.5&    0.220314& 0.220284& 0.220255& 0.220225& 0.220195 \\  \hline
		0.7&    0.30844& 0.308398& 0.308357& 0.308315& 0.308274 \\  \hline
		0.9&   	0.396565& 0.396512& 0.396458& 0.396405& 0.396352 \\  \hline
		1.1&   	0.484691& 0.484626& 0.48456& 0.484495& 0.48443 \\  \hline
	\end{tabular}
	\caption{The corrections in the deflection angle $\delta\alpha_D =\left.\alpha_D\right|_{\text{Kerr}}-\alpha_D$ for Sgr A* with $b=10^3M$, source star S2 at $r_s=1400M$, and varying $\Gamma$ and $a$; $\delta\alpha_D$ is in units of arcsec.}\label{T3}
\end{table}
\begin{table}[h!]
	\begin{tabular}{|c|c|c|c|c|c|}
		\hline	
		$\Gamma $  & $a=0.1$    & $a=0.3$     & $a=0.5$     &$ a=0.7 $   &$ a=0.9 $   \\
		\hline\hline
		0.1&	0.131438& 0.297011& 0.462519& 0.62796& 0.793335 \\  \hline
		0.3&    0.228683& 0.394244& 0.559738& 0.725166& 0.890528 \\  \hline
		0.5&    0.325928& 0.491476& 0.656957& 0.822372& 0.987721 \\  \hline
		0.7&    0.423174& 0.588708& 0.754176& 0.919578& 1.08491 \\  \hline
		0.9&   	0.520419& 0.68594& 0.851395& 1.01678& 1.18211 \\  \hline
		1.1&   	0.617664& 0.783172& 0.948614& 1.11399& 1.2793 \\  \hline
	\end{tabular}
\caption{The corrections in the deflection angle $\delta\alpha_D =\left.\alpha_D\right|_{\text{Schw}}-\alpha_D$ for Sgr A* with $b=10^3M$ and source at $r_s=10^5M$; $\delta\alpha_D$ is in units of arcsec. }\label{T2}
\end{table}

\begin{table}[h!]
	\begin{tabular}{|c|c|c|c|c|c|}
		\hline	
		$\Gamma $  & $a=0.1$    & $a=0.3$     & $a=0.5$     &$ a=0.7 $   &$ a=0.9 $   \\
		\hline\hline
		0.1&    0.114519& 0.255378& 0.396175& 0.536909& 0.677581 \\  \hline
		0.3&    0.202644& 0.343492& 0.484277& 0.624999& 0.76566 \\  \hline
		0.5&    0.29077& 0.431606& 0.572379& 0.713089& 0.853738 \\  \hline
		0.7&    0.378896& 0.519719& 0.66048& 0.801179& 0.941816 \\  \hline
		0.9&   	0.467021& 0.607833& 0.748582& 0.889269& 1.02989 \\  \hline
		1.1&   	0.555147& 0.695947& 0.836684& 0.977359& 1.11797 \\  \hline
	\end{tabular}
	\caption{The corrections in the deflection angle $\delta\alpha_D =\left.\alpha_D\right|_{\text{Schw}}-\alpha_D$ for Sgr A* with $b=10^3M$ and source star at $r_s=1400M$; $\delta\alpha_D$ is in units of arcsec. }\label{T4}
\end{table}

\begin{figure}
	\includegraphics[scale=0.7]{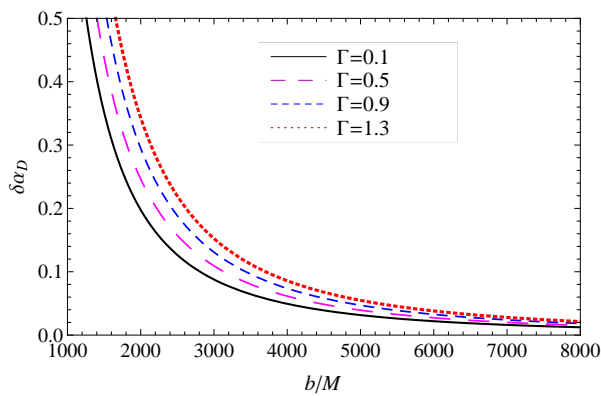}
	\caption{Correction in the deflection angle $\delta\alpha_D =\left.\alpha_D\right|_{\text{Schw}}-\alpha_D$ for rotating Kalb-Ramond black holes with $s=1$, $a=0.90$, and varying $b$; $\delta\alpha_D$ is in units of arcsec.}\label{DefAng}
\end{figure}
We aim to compare the lensing predictions of rotating Kalb-Ramond black holes (\ref{rotbhtr}) with those for Kerr and Schwarzschild black holes. We consider that the light coming from a distant source gets deflected by the Sgr A* black hole at the Galactic center ($M=4.0\times 10^6 M_{\odot}$, $r_o=d=8.3$ kpc) and reached the observer at Earth. In this case, the observer's distance $r_o$ is much larger than the impact parameter of light, whereas a source star may live in the bulge of our Galaxy. Therefore, even though the source can be still in the weak-field regime, we have to take account of finite-distance corrections due to the source. We calculate the light deflection angle and estimate the corrections from the Schwarzschild ($a=\Gamma=s=0$) and the Kerr ($s=0$) black holes. Tables \ref{T1} and \ref{T2} summarize these corrections, respectively, $\delta\alpha_D=\left.\alpha_D\right|_{\text{Kerr}}-\alpha_D$ and $\delta\alpha_D=\left.\alpha_D\right|_{\text{Schw}}-\alpha_D$, in the deflection angle for various values of black hole parameters and $r_o=\infty$ and $r_s=10^5M$. We consider the S2 star as the source, which in May 2018 approached the closest distance to Sgr A*, $r_s=1400\,M$, and presented the corrections in deflection angle in Tables \ref{T3} and \ref{T4}. One can see from Tables \ref{T1} and \ref{T2} that the presence of the Kalb-Ramond field significantly lowers the deflection angle as compared to those for the Schwarzschild or Kerr black holes, and the order of correction is arcsec which is within the resolution of today's observational facilities. The Kalb-Ramond field parameter $\Gamma$ gets nontrivially coupled with the black hole spin parameter $a$, such that, for fixed values of $\Gamma$ and $b$, the correction in the deflection angle from the Kerr black hole decreases with increasing $a$ (cf. Table \ref{T1}). Nevertheless, $\alpha_D$ increases with increasing $\Gamma$, and a non-rotating black hole causes a larger deflection angle as compared to the rotating one. This is because, in rotating spacetimes, the local inertial frame dragged along the black hole rotation, and it takes a shorter time for a prograde light ray to feel the gravitational pull. In Fig.~\ref{DefAng}, we have shown how the difference in the deflection angle $\delta\alpha_D=\left.\alpha_D\right|_{\text{Schw}}-\alpha_D$ varies with dimensionless impact parameter $b/M$ for different values of $\Gamma$. As the impact parameter $b$ increases, the correction in the deflection angle due to the Kalb-Ramond field get subsides.

\section{Conclusion}\label{sec6}
The detection of the Kalb-Ramond field, which appears as closed string excitations in the heterotic string spectrum, may provide profound insights to our understanding of the current Universe \cite{Kalb:1974yc,Chakraborty:2016lxo,Kar:2002xa}. The gravitational action, when nonminimally coupled to the Kalb-Ramond field, admits spherically symmetric hairy black holes \cite{Bessa:2019pom}. We derived the rotating counterpart of this solution, i.e., a rotating Kalb-Ramond black hole. The derived Kerr-like black hole has an additional Kalb-Ramond parameter $s$ besides mass $M$ and spin parameter $a$. The Kalb-Ramond field produces a hair that changes the structure of the rotating black hole through an extra term in the metric (\ref{rotbhtr}). Obviously, this rotating Kalb-Ramond black hole metric is asymptotically flat and encompasses Kerr ($s=0$), Kerr$-$Newman ($s=1$), Reissner$-$Nordstrom ($s=1,a=0$), and Schwarzschild ($s=0,a=0$) black holes. The rotating Kalb-Ramond black hole, like the Kerr black hole, still admits the Cauchy and event horizons, as well as the SLS. However, the radii of horizons and SLS decrease due to $s$, and the ergosphere is also affected and, thereby, can have interesting consequences on the astrophysical Penrose process.

Despite the complicated rotating metric (\ref{rotbhtr}), using the Komar prescription, we analytically derived the exact expressions for conserved mass $M_{\text{eff}}$ and angular momentum $J_{\text{eff}}$, valid at any radial distance. Furthermore, the presence of the Kalb-Ramond field significantly altered these conserved quantities as compared to those for the Kerr black hole, which is restored in the limit $s=0$; $M_{\text{eff}}$ and $J_{\text{eff}}$ decrease with increasing $\Gamma$ or $s$ for fixed values of other parameters. Nevertheless, the effect of the Kalb-Ramond field subsides at far distances from the horizon, as at asymptotically large $r$ ($r\to \infty$)  $M_{\text{eff}}$ and $J_{\text{eff}}$ take the values for the Kerr black hole. We further calculate the conserved quantity attributing to the generator of the event horizon to derive an interesting and important feature of the rotating black hole, namely, the generalized Smarr formula. 

Considering the observer and the luminous source at finite distances from the black hole, the analytical expression for the deflection angle in the weak-field limit is deduced, and also the higher-order correction terms to the deflection angle for the Schwarzschild and Kerr black holes due to the Kalb-Ramond field are calculated. We illustrated that the presence of the Kalb-Ramond field leads to a smaller deflection angle as compared to Kerr and Schwarzschild black hole values. This change in the deflection angle, for the supermassive black hole Sgr A* and the light source star in the bulge of the Galaxy, is as large as a few arcsec and, thus, feasibly measurable with present-day astronomical observations. For fixed values of black hole parameters ($M, \Gamma, s$) and impact parameter $b$, nonrotating Kalb-Ramond black holes are found to cause a larger deflection angle in contrast to rotating black holes. 

We also discussed the effects of the Kalb-Ramond field on black hole shadows, an extreme case of light gravitational lensing. It is found that shadows of rotating Kalb-Ramond black holes become smaller and more distorted with increasing field parameter $s$. The shadow observables, namely, area $A$ and oblateness $D$, are used to characterize the size and shape of the shadows and, thus, in turn, to extract the values of black hole parameters. The recent shadow observational results of the M87* black hole are used to put constraints on the Kalb-Ramond field parameter in the supermassive black hole context. 

More severe constraints can be expected by taking into account the surrounding accretion disk. The study of energy extraction and the particle production rate in the rotating Kalb-Ramond black hole spacetime are being considered for future projects. It will also be interesting to investigate the stability of the obtained rotating solution against the scalar perturbations in the context of gravitational wave observational data.

\begin{acknowledgements} S.G.G. and R.K. would like to dedicate this paper to Prof. M. Sami on his 65th birthday. S. G. G. thanks Department of Science and Technology INDO-SA bilateral project DST/INT/South Africa/P-06/2016, Science and Engineering Research Board for the ASEAN project IMRC/AISTDF/CRD/2018/000042, and also Inter University Centre for Astronomy and Astrophysics, Pune for the hospitality while this work was being done. R. K. thanks University Grant Commission for providing Senior Research Fellowship. A. W. thanks the staff at Zhejiang University of Technology for its hospitality, at which part of the work was done, and also thanks the National Natural Science Foundation of China for partial support with Grant No. 11375153 and No. 11675145. The authors also thank the anonymous reviewer for providing insightful comments which helped to improve the paper. 
\end{acknowledgements}

\noindent
\end{document}